\documentclass[aps,pra,twocolumn]{revtex4}
\usepackage{epsfig,amssymb,amsmath,mathrsfs,color,graphicx,float}
 \def\map#1{\mathcal #1}
\def\d{\operatorname{d}}\def\<{\langle}\def\>{\rangle}
\def\Tr{\operatorname{Tr}}\def\:{\hbox{\bf
    :}}

\def\grp#1{\mathsf{#1}}

\def\spc#1{\mathscr{#1}}

\def\Spec{\mathsf{Spec}}

\DeclareMathOperator\diag{diag}

\newtheorem{lemma}{Lemma}
\newtheorem{theo}{Theorem}
\begin{document}
\title{
Quantum replication
at  the Heisenberg limit
}

\author{Giulio Chiribella}
\email{gchiribella@mail.tsinghua.edu.cn}
\author{Yuxiang Yang}
\affiliation{Center for Quantum Information, Institute for Interdisciplinary Information Sciences, Tsinghua University, Beijing, 100084, China} 
\author{Andrew Chi-Chih Yao}
\affiliation{Institute for Interdisciplinary Information Sciences, Tsinghua University, Beijing, 100084, China} 
\homepage{http://iiis.tsinghua.edu.cn}
 
 \begin{abstract} 
No process in nature can perfectly clone an arbitrary quantum state.
But is it possible to engineer  processes  that replicate  quantum information with vanishingly  small error?  
Here we  demonstrate the possibility of probabilistic super-replication phenomena where $N$  equally prepared quantum clocks are transformed into a much larger number of $M$ nearly perfect replicas,  with  an error that rapidly vanishes  whenever $M$ is small compared to $N^2$. The quadratic replication rate is the ultimate limit imposed by Quantum Mechanics  to the proliferation of  information and is fundamentally linked with  the Heisenberg limit of quantum metrology.

  \end{abstract}
\maketitle

\section{Introduction}
No physical process can copy arbitrary quantum states on demand \cite{wootters82,dieks82}:   if such a process existed, we could build a device that distinguishes quantum states with arbitrary precision, violating the uncertainty principle and enabling faster-than-light communication \cite{herbert82}.  
Probabilistic processes like stimulated emission, however, seem to evade this restriction.   
In Einstein's treatment of stimulated emission \cite{einstein}, 
an excited atom interacting with a polarized photon is expected to produce sometimes a  second photon with the same polarization, effectively delivering a perfect clone. True that some other times the atom will spontaneously emit a photon of random polarization, but still, when stimulated emission occurs, a perfect clone has been produced.       
If we had on our side a quantum version of Maxwell demon, who separates the photons produced by stimulated emission from those produced by spontaneous emission,  we would be able to generate any desired number of clones with  a  non-zero probability.   
Unfortunately,  our imaginary helper is not allowed by the laws of Quantum Mechanics:  even with tiny probability, knowing whether or not stimulated emission took place  would lead to a  violation of the no-signalling principle (incidentally, this observation should be taken as a reminder that Einstein's treatment is just an approximation, whereas in  the actual quantum dynamics  stimulated and spontaneous emission happen in a coherent superposition).
However, nothing  forbids that other probabilistic processes, akin to spontaneous emission, could proliferate quantum information  beyond any previously conceived limit.  This possibility raises new fundamental questions:
Is it possible to engineer a process that duplicates a beam of $N$ equally prepared particles, producing a beam of $2N$ almost perfect clones?  What is the ultimate rate at which quantum information proliferate without significant errors?

Here we answer both questions: Although it is impossible to duplicate arbitrary quantum states, we devise a probabilistic mechanism that transforms an input beam of $N$ particles, equally prepared in a state $|\psi_t\>  =   e^{-it  H}  |\psi\>$ generated by  time evolution, into an output beam of magnified intensity, consisting of an overwhelming  number  $M $ of nearly perfect clones with a small  error that vanishes  rapidly whenever $M$ is small compared to $N^2$.      
 We name  this new  phenomenon  super-replication and show that it is intrinsically probabilistic, by proving that deterministic processes  can only produce  a negligible number of nearly perfect replicas. For example, for 100 linearly polarized photons, super-replication allows one to produce  $1000$ replicas with fidelity $99.9\%$, whereas the best deterministic process can only achieve fidelity $57\%$.
In addition, we show that no physical process, deterministic or not, can proliferate  quantum information at a rate larger than quadratic: any attempt to replicate quantum information beyond this limit is doomed to produce a joint output that has vanishing fidelity with the desired state.   To explain the roots of this fundamental limitation, we establish a deep link between quantum cloning and the precision limits of quantum metrology \cite{wineland,holland,caves,giovannetti04,giovannetti11,huelga,davidovich11}, showing that  the Heisenberg Limit  sets the ultimate bound to the replication rate of probabilistic processes, while the Standard Quantum Limit sets the corresponding bound  in the deterministic regime.  

\section{Results}

{\bf The Standard Quantum Limit for information replication.} 
 Optimal  cloning is a fundamental primitive in Quantum Information \cite{buzekhillery96,scarani05,cerffiurasek06}.  Its goal is to transform $N$ input copies of a quantum state  $|\psi_x\>$, randomly drawn from a set   $\{|\psi_x\>  \}_{x\in\mathsf X}$, into $M$ approximate copies that are as faithful as possible. The simplest figure of merit here is the fidelity  $F_{N\to M}$ between the $M$-particle state produced by the process and $M$ exact copies of the desired state, evaluated on the worst-case input state.  Inspired by information theory, we  now consider a sequence  of cloning processes that  transform $N$ input copies into $M   =  M(N)$ approximate copies and we say that the replication is reliable if the replication error vanishes in the limit of large $N$. As an error measure, we use  the infidelity $e_{N\to M}  :=  1-  F_{N\to M}$.  Here the crucial question is: How large can $M$ grow as a function of $N$ in a reliable replication process?        
To answer the question, we  introduce the notion of replication rate, saying that a replication process has rate $\alpha$ if   the number of extra-copies scales like $N^{\alpha}$. We say that a rate is achievable if there exists a sequence of reliable replication processes with that rate.  
We will now see that the achievable replication rates are determined by the precision limits of quantum metrology.
Our first key result in this direction is a  Standard Quantum Limit (SQL) for Quantum Replication: {No deterministic process can reliably replicate a continuous set of quantum states   at a rate larger than 1}.    
In other words, deterministic processes can only embezzle from nature  a negligible number of extra-copies.  

The derivation of the SQL for quantum replication  is based on the SQL for quantum metrology \cite{SQL}, applied to an arbitrary curve of states $\{ |\psi_t\>~|~  t \in  (-\epsilon,\epsilon) \}$ contained in the set of states that we are trying to clone. A sketch of proof is as follows:           The SQL  states that the variance in the estimation of $t$ from $N$ copies is lower bounded by 
$V_{N,  t}  \ge    \frac {c} N$, 
for a constant $c$ that can be set to $c=1$  with a suitable choice of parametrization.     Now, suppose that there is a sequence of reliable deterministic processes with rate $\alpha$.  The $N$-th process will produce an $M$-particle output state $\rho^{\text{out}}_{M,t}$,  where $M  \ge   N+ a  N^{\alpha}$ for some constant $a>0$, approaching the ideal target  $|\psi_t\>^{\otimes M}$ in the large $N$ limit.     For simplicity, let us assume  that the trace distance between $\rho^{\text{out}}_{M,t}$ and $|\psi_t\>^{\otimes M}$ vanishes as    $O(M^{-\beta})$ for some exponent $\beta  \ge 1$ 
(for sets of states of the form $\{|\psi_t\>  =  e^{-itH} |\psi\> ~  t\in \mathbb R\}$, this assumption is lifted in Supplementary Note 1).    Now, if two states are close, so is the variance in the estimation of $t$ from these two states: for every fixed estimation strategy, we have the bound  $    V^{\text{out}}_{M,t}    \le      V_{M ,t} +    \gamma M^{-\beta} $, where $V^{\text{out}}_{M,t} $ is the variance in the estimation of $t$  from $\rho^{\text{out}}_{M,t}$, $V_{M,t}  $ is the variance in the estimation of $t$ from $M$ copies, and $\gamma>0$ is a suitable constant.  However, applying a deterministic transformation  cannot reduce the variance of the optimal estimation strategy, denoted by $ V^{\min}_{N,t} $.  Hence, we have the bound  $  V^{\min}_{N,t}   \le V^{\text{out}}_{M,t}   \le      V_{M ,t} +    \gamma M^{-\beta} $. Choosing the best estimation strategy, we then obtain  
$V^{\min}_{N,t}  \le    V_{M,t}^{\min}   +  \gamma  M^{-\beta}$.   
Now, since the SQL is asymptotically achievable by a suitable strategy (cf. Methods),  for large $N$ the bound becomes  
$1/N  \le  1/{ (N +  a N^\alpha) }  + \gamma (N  + a N^\alpha)^{-\beta}$, up to terms that are negligible with respect to $N^{-1}$.
Clearly, this implies $\alpha  \le 1$, because otherwise we would have a contradiction for large $N$.   This establishes the SQL  for quantum replication.   

As a byproduct of the derivation, we also have that no deterministic process can replicate information with error vanishing faster than $N^{-4}$.  Indeed, by Taylor-expanding in the r.h.s. of the  inequality $1/N  \le  1/{ (N + a N^\alpha) }  + \gamma (N  + a N^\alpha)^{-\beta}$  we obtain the condition $\beta  \le 2 -\alpha$, which implies $\beta  \le 2$.  Using the relation between fidelity and trace-distance \cite{nielsen}, this means that the replication error is lower bounded as  $e_{N\to M}\ge O(N^{-4})$. In other words, for a reliable replication process the error cannot vanish faster than a low degree polynomial.     

The SQL limits not only deterministic replication, but also some instances of probabilistic replication: for example, probabilistic cloning has no advantage for arbitrary pure states  (this fact was observed in Ref. \cite{fiurasek} for $1$-to-$M$ cloning, but the argument can be easily extended to $N$-to-$M$ cloning).     More generally, it is easy to see that if the set of states $\{|\psi_x\>\}$ has strong symmetries, probabilistic processes do not lead to any improvement (cf. Methods).  A sketch of the argument is the following:  Any probabilistic process can be decomposed into the application of a filter---that transforms an input state $|\psi_x \>$  into the output state  $  |\phi_x\>  =  M_{\text{yes}} |\psi_x\> /  \|    M_{\text{yes}} |\phi_x\>\|$ for some suitable operator $M_{\text{yes}}$---followed by a deterministic process.   Now, if the set of states has strong symmetries, the optimal operator  $M_{\text{yes}}$ is forced to be equal to the identity, and the probabilistic process becomes equivalent to a deterministic one. This is the case for cloning of arbitrary pure states \cite{werner98},  coherent states \cite{cerf}, or  coherent spin states \cite{demkowicz04}, where the replication rates are bound to satisfy the SQL.   
\medskip 

\noindent {\bf The Heisenberg limit for  information replication.}  
We now restrict our attention to the replication of states of the form  $  |\psi_t\>  =  e^{-it  H} |\psi\>$,  $t\in\mathbb R$, where $H = H^{\dag}$ is a suitable Hamiltonian.  We call these states  clock states, as they can be generated through a time evolution obeying the Schr\"odinger equation.     
For simplicity, we focus on finite-dimensional quantum systems and we ignore the uninteresting case where $H$ is a multiple of the identity.   The main result here  is a Heisenberg limit for the probabilistic replication of clock states.  To get the result, we first establish a precision limit for probabilistic metrology,  where one is allowed to take advantage of filters  \cite{fiurasek06,gendra12}.  Our strategy is to apply the quantum Cram\'er-Rao bound \cite{helstrom,caves,QFImetric} to the states $|\phi_t\>=M_{\text{yes}} |\psi_t\> / \| M_{\text{yes}} |\psi_t\>\|$ emerging from the filter.  By explicit calculation (see Supplementary Note 2), we bound the variance at a given point $t$ as
\begin{align}\label{pcr}
V^{\text{prob}}_{t}  \ge  \frac 1  {  4\left(  \<  \phi_t  |  K_{t}^\dag K_{t} |\phi_t\>  -    \< \phi_t|  K_t  |\phi_t\>^2\right) }  \, ,
\end{align}
where  $K_t   = M_{\text{yes}}   H   M_{\text{yes}}^{-1}+       \frac{\<\psi_t|[H,M_{\text{yes}}^\dagger M_{\text{yes}}]|\psi_t\>}{2\|M_{\text{yes}}|\psi_t\>\|^2} $.
This innocent-looking  application of the Cram\'er-Rao bound leads immediately to a surprise: optimizing over all filters $M_{\text{yes}}$, the r.h.s. of equation (\ref{pcr}) can be made arbitrarily small, suggesting the possibility of unlimited precision (cf.  Methods).  Note however, that in order to attain the equality in equation (\ref{pcr}), one should adapt the choice of filter to the value of $t$, the unknown parameter that one is trying to estimate.  In practice, this is not a realistic scenario.   In the case of periodic evolution,   a more realistic setting  is to have  $t$  distributed according to a uniform prior $p(t)$ over the period.  In this case, the precision that can be achieved on average is still limited:  denoting by $V_p^{\text{prob}}$ the expected variance  when the particles pass the filter,  in Supplementary Note 3 we prove the bound
\begin{align}\label{protohl}
V_{p}^{\text{prob}}  \ge  \frac 1  { (  E_{\max}  - E_{\min}  )^2  } \,  ,
\end{align}  
where $E_{\max}$  ($E_{\min}$) is the maximum   (minimum)  eigenvalue of the energy such that $  |\psi\>$ has non-zero overlap with the  corresponding eigenspace. We note that $  (E_{\max} - E_{\min})^2/4$  is equal to the maximum of the variance of $H$ over all possible states $|\phi\>$ contained in the subspace  generated by the input states $\{  |\psi_t\> \}$.  Such maximum is achieved by the ``NOON state"  $ |\phi\>  =  (|E_{\max}  \>  +  |E_{\min}\>)/\sqrt 2 $, where $  |E_{\max}\>$  ($|E_{\min}\>$) is an eigenstate of $H$ corresponding to the eigenvalues $  E_{\max}$ ($E_{\min}$). Since these states  are the best states for deterministic metrology  \cite{metroprl},  our result implies that  the best strategy for probabilistic metrology with uniform prior consists just in using a filter that generates the NOON state, and then applying the optimal deterministic estimation strategy. 
Up to a small correction, the same result of equation (\ref{protohl}) can be obtained when the evolution is non-periodic, by approximating the uniform distribution with a Gaussian with large variance  (cf. Supplementary Note 4).  

When $N$ identical copies  of $|\psi_t\>$ are available,  equation (\ref{protohl}) gives the  Heisenberg scaling
\begin{align}\label{hl}
  V_{p, N}^{\text{prob}}  \ge  \frac 1  { N^2 (  E_{\max}  - E_{\min}  )^2}  \,.      
 \end{align}
 Equipped with this bound, we can now derive the Heisenberg Limit (HL) for Quantum Replication: 
{No physical process can reliably replicate a set of clock states at a rate  larger than $2$}.    Here is a sketch of proof:   The filter transforms  the product state $|\psi_t\>^{\otimes N}$ into the entangled state $M_{\text{yes}} |\psi_t\>^{\otimes N}/  \|  M_{\text{yes}} |\psi_t\>^{\otimes N} \| $.   For this state, equation (\ref{hl}) gives  the bound $V^{\text{prob}}_{p, N}  \ge c/N^2 $ with $c=  (E_{\max}  -  E_{\min})^{-2}+\epsilon$ ($\epsilon=0$ in the case of periodic evolution). Now, suppose that there exists a sequence of  probabilistic processes with replication rate $\alpha>1$, and suppose that the $N$-th process in the sequence produces  $M  \ge   N  +  a N^{\alpha} $ approximate copies, with a  trace distance from the ideal target vanishing  asymptotically as $O(1/M^{\beta}),  ~ \beta  \ge 1$ (this assumption is lifted in Supplementary Note 5). Then, the variance of the estimation from the output state $\rho^{\text{out}}_{M,t}$---denoted by $V_{M,t}^{\text{out}}$---will be close to the variance of the estimation from $ |\psi_t\>^{\otimes M}$:  for every $t$, we will have $  V^{\text{out}}_{M,t} \le   V_{M,t}  +  \gamma/M^{\beta}$, for some constant $\gamma$ independent of $t$.   
Taking the average variance over the uniform prior $p(t)$,   we obtain the relation $ V^{\text{out}}_{M,p}  \le  V_{M,p}  +  \gamma/M^{\beta}$, where $V_{M,p}^{\text{out}}$ and $V_{M,p}$ are the averages of $V_{M,t}^{\text{out}}$ and $V_{M,t}$, respectively.  
  By definition, the average variance $V_{M,p}$ is lower bounded by the SQL and the bound is attained  in the large $M$ limit by choosing a suitable measurement (cf. Methods).     Hence, by Taylor expanding the terms in $M$ and keeping the leading order terms, we obtain  
$c N^{-2}  \le  4  N^{-\alpha}   +  O  (N^{-\alpha })$, 
which implies $  \alpha  \le 2$.  Note that  here $\beta$ disappeared from the equation:  as there is no upper bound on $\beta$, in principle the error can vanish faster than any polynomial!     

The HL leaves lots of room for  replicating quantum information:   for every  rate $1<\alpha <2$ one has the chance not only to duplicate the input copies, but also to produce an overwhelming number of replicas, with an error that vanish faster than any polynomial.  In the next paragraph, we will exhibit explicit protocols that have all these features.   
Our protocols are necessarily probabilistic, as a consequence of the SQL, which constrains deterministic processes to  produce a negligible number of extra-replicas. 

\medskip

\noindent {\bf   Probabilistic super-replication.}  Here we show that  clock states can be reliably replicated at any rate allowed by the HL.  When the rate is larger than (or equal to)  1, we call the process  super-replication to emphasize the fact that it beats the SQL.  The key idea to achieve super-replication is to devise a filter that modulates the Fourier amplitudes of the  wavefunction in a way that enhances the replication performances to the maximum rate allowed by Quantum Mechanics.            To move to the Fourier picture, we express   the input states as  $|\psi_t\>  =   \sum_{E  \in  \Spec (H)}    \sqrt{p_E}   e^{-it   E}  |E\>$, where $\Spec (H)$ denotes the spectrum of the Hamiltonian, $   |E\>$ is an eigenvector of $H$ with eigenvalue  $ E$, and $p_E$ is the probability that a measurement of the energy gives outcome $E$. 
When $N$ identical copies are given, the joint state can be expressed as 
\begin{align}\label{state}
|\psi_t\> ^{\otimes N}   =    \sum_{ E \in \Spec (H^{(N)})}      e^{ -i  t  E}       \sqrt{ p_{ N, E}  }  | N, E  \>     ,  
\end{align} 
where $H^{(N)}  =  \sum_{i=1}^N   H_i$ is the total  Hamiltonian ($H_i$ denoting the operator $H$ acting on  the $i$-th system), $|N,E\>$ is an eigenvector of $H^{(N)}$ for the eigenvalue $E$, and $p_{N,E}$ is the probability that a measurement of the total energy gives outcome $E$.    Choosing a filter $M_{\text{yes}}$ that is diagonal in the energy eigenbasis, we can modulate the Fourier amplitudes of the  state   (\ref{state}) in any way we like.     Of course, since we aim at producing $M$ perfect copies, the natural choice is  to replace $N$ with $M$ in the probability distribution $p_{N,E}$.   However,  the spectrum of $H^{(N)}$ may be not be contained in the spectrum of $H^{(M)}$ and in general one needs to shift the energy values by a suitable amount $\delta E_0  \approx   (M -  N)  \< \psi|  H|\psi \>$  (see Supplementary Note 6). 
With this choice,    the state after the filter is  projected to the entangled state
$\left|\Phi^N_t\right\>  \propto        \sum_{E \in \Spec (H^{\otimes N})}      e^{ -i  t E }       \sqrt{ p_{ M, E+ \delta E_0} }   | N,  E     \>  $. 


The state  $|\Phi^N_t\>$ will now act as a quantum program, containing the instructions that will be used by a deterministic quantum device to generate $M$ approximate clones.        
For this purpose, we use a  device that coherently transforms each  eigenstate $|N,E \>$   into the corresponding eigenstate $|M, E + \delta E_0 \>$, thus producing the state    $\left|\Psi^M_t\right\>  \propto        \sum_{E  \in \Spec(H^{(N) })}      e^{- i  t E}       \sqrt{ p_{ M,  E +  \delta E_0 }   } | M, E + \delta E_0   \>  $.   It is not hard to see that  the  state  $|\Psi^M_t\>$ has high fidelity with the desired state $|\psi_t\> ^{\otimes M} $ for every replication rate allowed by the Heisenberg limit.  Precisely, for large enough $N$ one has the bound 
\begin{align}\label{fidelity}
F_{N\to M} \ge      1  -   2  K   \exp  \left(-   \frac{ 2 p^2_{\min}   N^2}{  M   }  +  \frac{4 N}{M K}\right)  ,
\end{align}
where $K$ is the number of energy levels $E$ such that $  p_E \not =  0$ and  $p_{\min} = \min\{ p_E :  p_E  \not = 0\}  $ (see Supplementary Note 7  for the proof).
   Equation (\ref{fidelity}) shows that the fidelity approaches 1 faster than any polynomial whenever $M$ is of order $N^{\alpha}$ with $\alpha < 2$, that is, whenever the replication rate satisfies the HL.   
  A strong converse can be proven:  every process replicating quantum states at a rate higher than $ 2$ must have  vanishing fidelity in the limit of large $N$, as  showed in Supplementary Note 5.

Thanks to our filter, we have been able to embezzle from nature a large number of replicas.  
The improvement is striking if one compares  it   with the performances of  standard cloning processes.   Let us illustrate this fact for the replication of linear polarization states $|\psi_t\>  =  \cos t  |V\>  +   \sin t |H\>$.    
In this case,  the best deterministic process is the phase-covariant cloner of Ref. \cite{dariano},  and its fidelity is    $F^{\text{det}}_{N \to M}  \approx  \frac{2\sqrt{MN}}{M+N}$ in the asymptotic limit of large $N$ and $M$.   In agreement with our SQL, the fidelity  vanishes whenever $M$ is of order $  N^{ 1 +\epsilon}$, $\epsilon> 0$:  As we proved in  general, deterministic  processes can only produce a number of extra replicas that is a negligible fraction of the number of input photons.   In stark contrast, for $N= 100$ input photons, our filter can produce $M=1000$ approximate copies with fidelity  $F_{\text{prob}}  = 0.9986$, whereas the fidelity of  the optimal deterministic  cloner  is only  $F_{\text{det}}  = 0.5739$. 
In the example of linearly polarized photons our filter is provenly optimal, as it coincides with the optimal probabilistic cloner of Ref. \cite{fiurasek}. Quite surprisingly,  the possibility of super-replication was not recognized in Ref. \cite{fiurasek}, where  the advantage of probabilistic processing was conjectured to be only of the order of one percent.  

\begin{figure}
\centering
\includegraphics[totalheight=80mm]{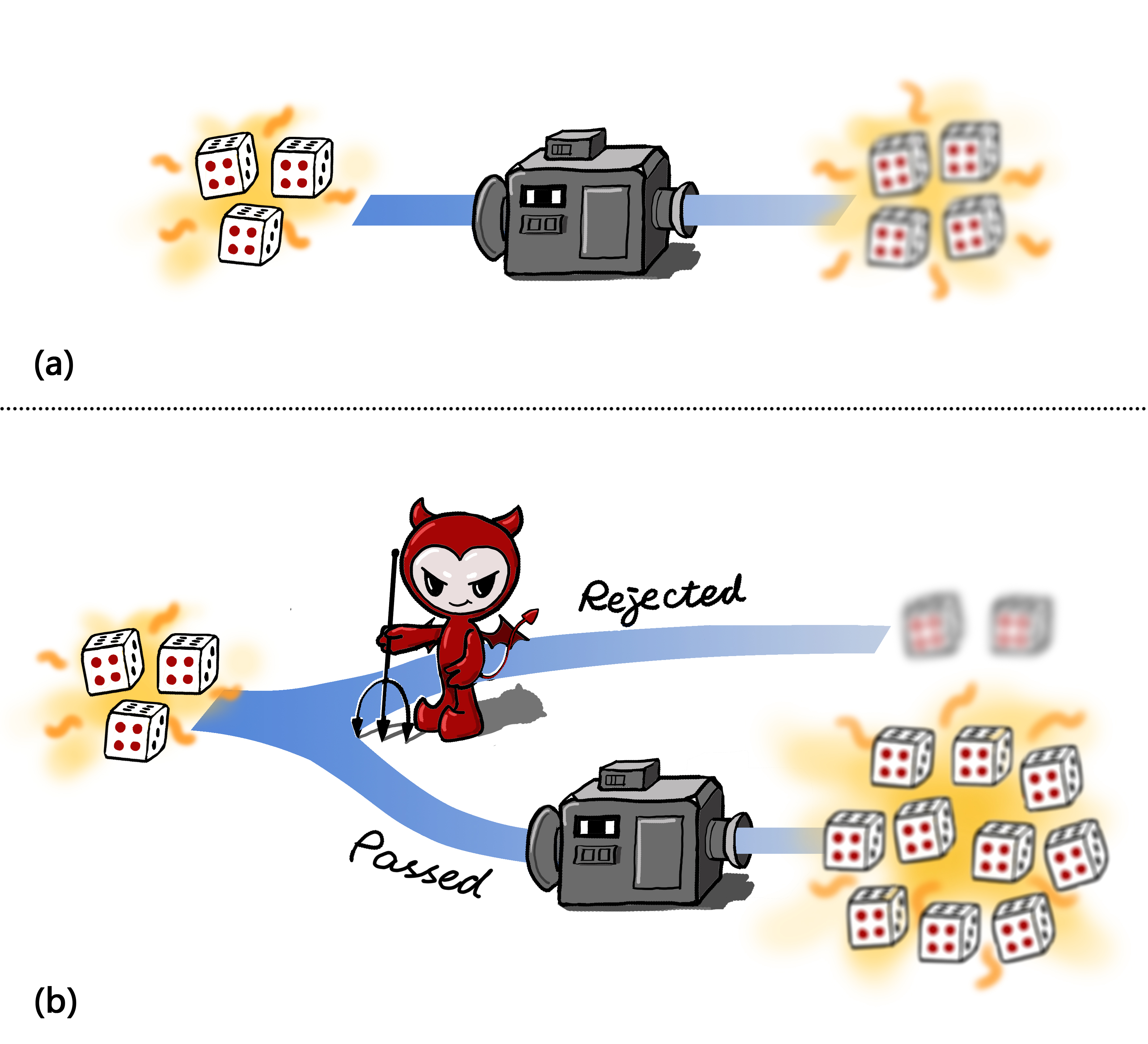}
\caption{{\bf The advantages of probabilistic super-replication.}   ({\bf a})  Every deterministic process is constrained by the Standard Quantum Limit:   the number of clones that can be produced reliably from $N$ input copies is of order $N$ and the fidelity of the clones cannot approach 1  faster than the inverse of a polynomial of degree  4.   ({\bf b})   Super-replication at the Heisenberg limit:  the  cloning performances  can be dramatically increased by a probabilistic filter, depicted here as a ``quantum Maxwell demon" that separates two branches of the wavefunction.   In the successful branch, any number of clones of order  $N^{2-\epsilon}$  can be produced with fidelity approaching 1 faster than any polynomial.   
}\label{fig:demon}
\end{figure}

\begin{figure}
\centering
\includegraphics[totalheight=60mm]{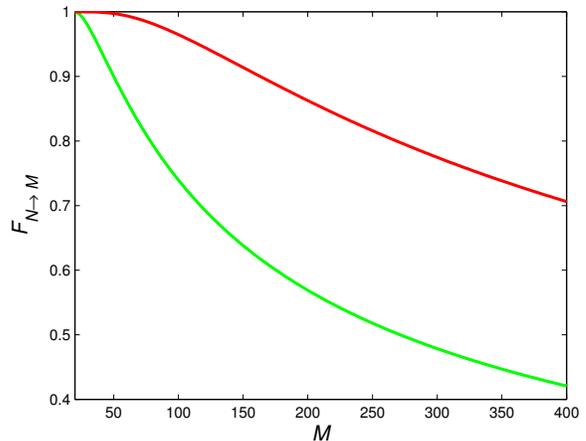}
\caption{ {\bf Replication of linearly polarized photons.}  The fidelity of the best cloning processes  for linearly polarized photons $|\psi_t\>  =  \cos t |V\>  +  \sin t |H\>$ is plotted here for $N=20$ input copies, with the number  $M$ of output clones ranging from $20$ to $400$. The  red (green) line refers to optimal cloning with (without) filter.  }\label{fig:fidelity}
\end{figure}

The different features of replication processes at the HL and at the SQL are illustrated in Fig. \ref{fig:demon} in the case of linearly polarized photons. 
The advantage of the filter is clear also in the non-asymptotic setting:  a plot comparing the performances of replication with and without the filter in the case $N=20$ is presented in Fig. \ref{fig:fidelity}.  
What makes the improvement even more dramatic is that the quality of the replicas is measured by the global  fidelity between the output state and the desired joint state  of $M$ perfect copies, which is much smaller than the fidelity that each single copy has with the state $|\psi_t\>$.    
\medskip 

\noindent{\bf Maximizing the probability of success.} The dazzling performances of super-replication come at a price:  the probability that the input systems pass the filter has to decay  with $N$. Indeed, in Supplementary Note 1  we already 
showed 
a strong converse of the SQL:  every deterministic process producing replicas at a rate higher than the SQL must have vanishing fidelity.

For our filter,  the probability of super-replication at rate $\alpha>1$ decreases exponentially fast as
$p_{\text{yes}} [N  \to M]\le         e^{-  k N} $ for a suitable constant $k>0$  independent of $\alpha$ (see Supplementary Note 8).
One may ask whether there are ways to achieve super-replication with a larger success probability.  It turns out that the answer is yes:  in Supplementary Note 9 we show that a process producing $M  \le  O( N) $ replicas can have a success probability $p_{\text{yes}} [N\to M]$ going to zero as $1/N^{\delta}$ for every desired $\delta>0$. 
Most importantly, for a super-replication process with rate  $\alpha =  1+\epsilon$, $\epsilon>0$, we show how to increase the success probability to  $p_{\text{yes}} [N \to M] \ge   e^{-N^\delta} $  for every desired exponent $ \delta  >  \epsilon$. 
However, no further improvement is possible below the critical value $\delta_{\rm c}  =  \epsilon$:  any  process with success probability scaling as $ p_{\text{yes}}  [N  \to  M]   =   a   e^{-b N^\delta}$,  $\delta  <  \epsilon$,  $a,b>0$  must have vanishing fidelity in the asymptotic limit, as observed in Supplementary Note 1. 

These results identify the optimal exponent for the decay of the success probability in a super-replication process, pinning down the tradeoff between the replication rate and success probability. Interestingly, the choice between the advantage of a high replication rate and that of a unit success probability can be always delayed to the very last moment: indeed, it is easy to see that asymptotically the best deterministic replication process is just given  by the coherent transformation $  |N,E \>  \mapsto  |M, E  +  \delta E_0\>$ that we used in our protocol. Hence, an alternative way to achieve super-replication is to apply first the best deterministic process and later to modulate the Fourier amplitudes of the wavefunction using a probabilistic filter.

\medskip

\noindent {\bf Many-worlds fairness. }
  Super-replication can be achieved not only for one-parameter families of clock states, but also for different manifolds  of states, including the manifold of all maximally entangled states of two identical systems. 
However, super-replication is not a generic feature. As we already mentioned, if one tries to copy an arbitrary---as opposed to linear---polarization state, then no filter is going to help: the  performances with filter are equal to the performances without filter.  More generally, for generic quantum systems,  no probabilistic filter can make an arbitrary quantum state more copiable.   This property is quite compelling when considered  from the  angle of the many-worlds  interpretation of Quantum Mechanics \cite{everett}, because it states that no-branch of the wavefunction of the universe offers an advantage over the others in  replicating the information contained in a completely unknown state.   
 Regarding different branches as ``different worlds",  we can formulate this as a fundamental principle, which we name many-worlds fairness: the maximum rate at which arbitrary information can proliferate is the same in all possible worlds.   Many world fairness rules out Quantum Mechanics on real Hilbert spaces  \cite{stueckelberg60}, an  alternative physical  theory where photons can have only linear polarizations.
 Thanks to this observation,  one can provide a new answer to the old question``Why are physical systems described by complex---instead of real--- Hilbert spaces?".   Traditionally, the standard answer has been to invoke  local tomography, the property that one can completely identify a mixed state from the statistics of local measurement on the components. 
However, one may consider this as an \emph{ad hoc} requirement  \cite{wootters12}, and, in fact,  there are even reasons to prefer real Quantum Mechanics to its complex version, as it was recently pointed out by  Wootters  \cite{wootterslast}.   Balancing this fact, many world fairness offers a new reason (other that the usual local tomography) in favour of  complex Quantum Mechanics.

\section{discussion}  

Super-replication has been introduced here from a theoretical point of view.  But is it possible to implement it experimentally?   Luckily,   
  very recently  there have been experimental breakthroughs on the closely related topic of probabilistic  amplification of coherent states of light \cite{ampli1,ampli2,ampli3,ampli4,ampli5}.  
 Although the translation to our case is not immediate, we suggest that super-replication of linearly polarized photons could be achieved through a suitable sequence of amplitude damping channels, which for the polarization play the role of the photon subtraction for coherent states.  An alternative approach is to first encode the state of the input copies  into a coherent state via matter-light teleportation \cite{lightmatter}, amplify the coherent state in a probabilistic fashion, and then teleport back.  This scheme  provides a new application of the existing experimental schemes for coherent state amplification, making them the building block for the replication of quantum information at the Heisenberg limit. Finally, a third avenue towards implementing super-replication would be through stimulated emission, combined with a suitable monitoring of the emitting atoms. Simulated emission has been discussed extensively in connection with deterministic cloning \cite{milonni,mandel,simon}, but its potential for implementing probabilistic processes is still  unexplored.        
     
In relation to the existing literature, it is interesting to comment on the relation between our results and previous works on probabilistic cloning and probabilistic estimation.   The idea that in some situations  even non-orthogonal states can be copied perfectly using a probabilistic device  was first introduced by Duan and Guo \cite{duan98}, who showed that a set of pure quantum states can be copied perfectly if and only if they are linearly independent.  This means that for a single photon one can perfectly copy two polarization states, while for $N$ photons one can copy at most $N+1$ states.   Subsequent works  \cite{fiurasek,rl,xie} showed that probabilistic devices can improve the performances of approximate cloning for continuous sets of states, leading to nearly ideal performances in the case of coherent states with fixed amplitude.   The nearly perfect cloning of coherent states may lead one to believe that in general there is no limit to the amount of clones that can be produced probabilistically. Contrarily to this intuition, we have proven here that for finite quantum systems the Heisenberg limit sets the ultimate bound $M= O(N^2)$ to the number of clones that can be produced reliably.  Moreover, we have shown that for arbitrary one-parameter families of states, any replication rate allowed by the Heisenberg Limit can be achieved and we identified exactly the exponent at which the probability of success has to decay.  

%
Finally, the relation between cloning and estimation has been extensively investigated in the literature \cite{gisin97,bruss98,bae06,chiribella06,chiribella11} in terms of single-copy fidelities.   However, none of the approaches proposed so far was suitable to derive limits on the asymptotic replication rates, nor to connect the latter with quantum metrology.   
Recently, it has been observed that the precision of quantum  estimation can be improved using probabilistic strategies \cite{fiurasek06,gendra12} where, depending on the outcome of the filter, one can decide to abstain from estimating the parameter.   The mechanism of abstention can sometimes boost the precision  from the SQL $1/N$  to the HL  $1/N^2$.   The connection between these results and super-replication  is made clear by our approach:  the fact that a probabilistic filter can improve estimation until the HL implies that $N^2$, rather than $N$,  is the upper bound for the replication  rate of  the states $|\psi_t\>  =  e^{-it H}  |\psi\>$.   However, there is an important catch:  the filter that achieves replication at the HL  is not the same filter that achieves estimation at the HL.  In fact, if we were to use the same filter needed in Refs. \cite{fiurasek06, gendra12}, we would not be able to reduce the  error down to zero.

\section{Methods}

\noindent{\bf Decomposition of quantum instruments.}
To describe the most general probabilistic processes allowed by Quantum Mechanics, we use   the framework of quantum instruments
 \cite{davies70,ozawa84,operationalphys}.   
A quantum instrument with input  (output) Hilbert space  $\spc H_{\text{in}}$  ($\spc H_{\text{out}}$)  is a collection of completely positive (CP), trace non-increasing linear maps $\{ \map P_j \}_{j \in \mathsf Y}$, where each map transforms  density matrices on $\spc H_{\text{in}}$ into (sub-normalized) density matrices on $\spc H_{\text{out}}$.  If the input system is prepared in the density matrix $\rho$, the probability that outcome $j$ is $p(j|  \rho)   =  \Tr [  \map P_j (\rho)]$.  Conditionally to the occurrence of outcome $j$, the state of the output system is $\rho_j^{\text{out}}  =   \map P_j (\rho)/\Tr[\map P_j (\rho)]$.   We note an elementary fact about quantum instruments:  any  quantum instrument can be decomposed into a pure measurement on the input system followed by quantum channel depending on the outcome.   
By ``pure measurement on the input system", we mean a quantum instrument $\{\map M_j\}_{j\in\mathsf  Y}$  of the special form  $\map M_j (\rho)  =   M_j \rho M_j^\dag$  for some operator $M_j$  on $\spc H_{\text{in}}$.   By ``quantum channel",  we mean a trace-preserving CP map $\map C_j$, transforming density matrices on $\spc H_{\text{in}} $  into  (normalized)  density matrices on $\spc H_{\text{out}}$.   In formula, our claim is that every  CP map $\map P_j$ can be decomposed as \begin{align}\label{instrudecomp}
\map P_j  =    \map \map \map C_j  \map M_j \, . 
\end{align}  
The proof is as follows:   Let $\map P_j^\dag $ be the  adjoint of the map $\map P_j$, defined by  $ \Tr[ \map P_j^\dag  (A)   \rho   ]  =  \Tr[  A  \map P_j (\rho)]$
 for every operator $A$ and for every density matrix $   \rho$.   Define the operator 
 $M_j  :=  \sqrt  {  \map P_j^\dag  (I)}$
 and  the map $\map L_j$ by the relation $\map L_j (\rho) : =   \map P_j     ( M_j^{-1}   \rho  M_j^{-1})$, where $M_j^{-1}$ is the inverse of $M_j$ on its support.    If $M_j$ is invertible, then $\map C_j:  =  \map L_j$ is trace-preserving:  
 \begin{align*}
 \Tr[   \map L_j(  \rho) ]  & =  \Tr[  \map P_j    ( M_j^{-1}   \rho  M_j^{-1}) ]\\
 &  =     \Tr\left\{ \map P^\dag_j  (I)        \left[ \map P^\dag_j  (I)  \right]^{-1/2}   \rho     \left[ \map P^\dag_j  (I)  \right]^{-1/2}    \right\}\\ 
 &  =  \Tr[\rho].
 \end{align*}        In this case, by definition we have $  \map P_j  (\rho) =  \map C_j (M_j  \rho M_j^\dag)$ for every $\rho$.   
       If $M_j$ is not invertible,  we can define the trace-preserving map $\map C_j  (\rho)  : =  \map L_j  (\rho)  +  P_\perp  \rho P_\perp$, where $P_\perp$ is the projector on the kernel of $M_j$.  Again, the definition implies     $  \map P_j  (\rho) =  \map C_j (M_j  \rho M_j^\dag)$ for every state $ \rho$.   $\blacksquare$

\medskip 

For the purposes of this paper, it is enough to consider measurements with two outcomes $\{\text{yes}, \text{no}\}$, referred to as filters.  The filter induces a bifurcation of the wave-function and selects one particular branch, corresponding to the successful outcome yes. 
 
\medskip 

\noindent{\bf Symmetry constraints on probabilistic cloning.}  Consider a quantum cloning problem where the set of states $\{  |\psi_x\>\}$ has a group of symmetries, denoted by $\grp G$.  
This means that, for every unitary operator $U_g$ representing the action of a symmetry transformation in the group, one has $\{  U_g  |\psi_x\>\}_{x\in\mathsf X}  =  \{ |\psi_x\>\}_{x\in\mathsf X}$.   A probabilistic cloner is  described by a quantum instrument $\{  \map P_{\text{yes}},  \map P_{\text{no}}\}$ from the space of $N$ copies to that of $M$ copies, where the CP maps  $\map P_{\text{yes}}$ and $\map P_{\text{no}}$ correspond to the successful and unsuccessful instances, respectively.   
Conditional to the successful outcome, the fidelity of the output state with the desired $M$-copy state is  
\begin{align}\label{fid}
F^{\text{prob}}_x [ \map P_{\text{yes}}]  =    \frac{  \<  \psi_x|^{\otimes M }  ~  \map P_{\text{yes} }   \left[  (|  \psi_x\>\< \psi_x|  )^{\otimes N}   \right]     |  \psi_x\>^{\otimes M }        }{  \Tr \left\{ \map P_{\text{yes} }  \left [  (|  \psi_x\>\< \psi_x|  )^{\otimes N}   \right] \right \} } .  
\end{align}

The goal of cloning is to maximize the worst-case fidelity   $F^{\text{prob}}_{\text{wc}}  [  \map  P_{\text{yes}}] :  = \inf_{x }   F^{\text{prob}}_x  [  \map  P_{\text{yes}}]$. 
Due to the symmetry of the set of states, the maximization can be restricted without loss of generality to the set of covariant CP maps, i.e. of CP maps satisfying the relation  
\begin{align}\label{covariantP}
U_g^{\otimes M}  \left[ \map P_{\text{yes}}  (\rho) \right]  U_g^{\dag \otimes M } =    \map P_{\text{yes}}   \left(  U_g^{\otimes N} \rho  U_g^{\dag \otimes N } \right),  
\end{align} 
for every group element $g$ and for every quantum state $\rho$.    The proof is standard and we refer the interested reader to similar proofs provided in the literature, such as those of Refs. \cite{holevo,ozawa} for covariant measurements and that of Ref.\cite{extremeclon} for  covariant cloning channels.

Let us work out the implication of symmetry, starting  from the decomposition $\map P_{\text{yes}} =  \map C_{\text{yes}}  \map M_{\text{yes}}$ of equation (\ref{instrudecomp}).    It is easy to see that the covariance of $\map P_{\text{yes}}$ implies  the commutation relation $[ M_{\text{yes}}, U_g^{\otimes N}]  =  0$, for every group element $g$.  Indeed, using the definition $M_{\text{yes}} := \sqrt{  \map P_{\text{yes}} (I)}$ and the covariance of $\map P_{\text{yes}}^\dag$, we have 
\begin{align*}
U_g^{\otimes N}  M_{\text{yes}}  U_g^{\dag \otimes N}  &  =    U_g^{\otimes N}    \sqrt{  \map P_{\text{yes}}^\dag  (I)}  \, U_g^{\dag \otimes N}  \\ 
&  =  \sqrt{\map P_{\text{yes}}^\dag  \left(  U_g^{\otimes M}  U_g^{\dag \otimes M } \right)}  \\
&  = \sqrt{ \map P_{\text{yes}}^\dag  \left( I  \right)}  \\
 &  \equiv  M_{\text{yes}}. 
\end{align*}     Now, if the group of symmetries is sufficiently large, the action of the unitaries $U_g^{\otimes N}$ can be irreducible in the subspace containing the input states $|\psi_x\>^{\otimes N}$.  When this happens, the Schur's lemma imposes that the filter $M_{\text{yes}}$ be a multiple of the identity, and, therefore $\map P_{\text{yes}}  \propto \map C_{\text{yes}}$.   By equation (\ref{fid}), this means that the fidelity achieved by the CP map $\map P_{\text{yes}}$ is the same as the fidelity achieved by the quantum channel $\map C_{\text{yes}}$.   In summary, we have proven the following fact:        
 The maximum of the fidelity over all probabilistic cloners is equal to the maximum of the fidelity over deterministic cloners whenever the action of  the unitaries $\{U_g^{\otimes N}\}$ is irreducible in the subspace spanned by the input states  $\{  |\psi_x\>^{\otimes N}\}$.      This simple observation has several important consequences.   A first consequence, noted by Fiur\'a\v sek  \cite{fiurasek}, is that the probabilistic processes do not offer any advantage for $1$-to-$M$ cloning of arbitrary pure states.  Our approach allows to reach the same conclusion in a fairly broader range of cloning problems: in particular, it implies that probabilistic processes offer no advantage in the case of $N$-to-$M$ cloning of arbitrary pure states  \cite{werner98},  $N$-to-$M$ cloning of  coherent states \cite{cerf} and spin-coherent states \cite{demkowicz04},  $1$-to-$M$ cloning of two Fourier transformed bases \cite{extremeclon}  and  of all the sets of states considered in Ref. \cite{extremeclon},  and 1-to-$M$ phase-covariant cloning for qubit states on the equator of the Bloch sphere \cite{dariano}.

\medskip

\noindent{\bf Pointwise vs average Cram\'er-Rao bound.}    The Cram\'er-Rao bound (CRB) is the cornerstone of quantum metrology  \cite{wineland,holland,caves,giovannetti04,giovannetti11,davidovich11}. For a family of clock states $  |\psi_t\>  =  e^{-it H} |\psi\>$, it states that the variance in the estimation of $t$ from the state $  |\psi_t\>$ is lower bounded by  $V_t \ge  1/Q_t$, where $  Q_t   :=  4  (   \<  \psi_t |  H^2  |\psi_t\> -     \<  \psi_t |  H  |\psi_t\>^2  )$ is the Quantum Fisher Information (QFI).     The equality in the CRB   can be achieved if one adapts the estimation strategy to the value of $t$.    This can be done in an asymptotic scenario when one is given a large number of copies of $|\psi_t\>$ and uses a fraction of them to obtain a rough estimate of $t$ \cite{gillmassar}. However, if instead of having $N$ copies one has a single $N$-particle entangled state, the CRB may not give an achievable lower bound.  For example, Hayashi \cite{hayashi} showed that the CRB for phase estimation predicts the correct scaling $c/N^2$, but with a constant $c$ that is not achievable.     In the case of probabilistic metrology, the issue about the achievability of the CRB appears in a more dramatic way: if we allow one to adapt also the choice of the filter $M_{\text{yes}}$ to the value of $t$, then the CRB promises unlimited precision.     It is instructive to see how this phenomenon can happen.     For a given $t_0$, choose a parametrization where $  \<\psi_{t_0}|  H  |\psi_{t_0}\>=0$ (this can always be done by  redefining $H$ as $H'  =  H  - \<\psi_{t_0}|  H  |\psi_{t_0}\> $). Then,  define the vector 
\begin{align*}  
|\psi_{t_0}^\perp\>  :  =  \frac{  H  | \psi_{t_0}\> }{\|   H  | \psi_{t_0}\> \| },
\end{align*}
which, by construction satisfies $  \<  \psi_{t_0}^\perp  | \phi_{t_0} \>  =0$.  
Now, choose the filter operator  
\begin{align*} M_{\text{yes},t_0}  :=  \epsilon  |\psi_{t_0}\>\<  \psi_{t_0}|    +  | \psi^\perp_{t_0}\>\<  \psi^\perp_{t_0}|
 \end{align*} 
The filter does not change the state $|\psi_{t_0}\>$:
\begin{align*}
|\phi_{t_0} \>   =   \frac{  M_{\text{yes},t_0}  |\psi_{t_0} \> }{  \|  M_{\text{yes},t_0}  |\psi_{t_0}\>  \| }  \equiv  | \psi_{t_0}\> \, .
\end{align*}
However, it amplifies the time-derivative at $t_0$: 
\begin{align*} 
\left.   \frac{  i    \d |\phi_t\>}{ \d  t}      \right |_{t=  t_0}    & =    K_{t_0}  |\phi_{t_0}\> 
 \\  &=  \left(M_{\text{yes},t_0}   H   M_{\text{yes},t_0}^{-1} \right)   |\psi_{t_0}\>\\
&  =    \frac{ H |\Psi_{t_0}\>}{\epsilon    }\\
&  = \left.   \frac 1 \epsilon  \frac{  i    \d |\psi_t\>}{ \d  t}     \right |_{t=  t_0}   ,
\end{align*}
where in the second line we used the explicit expression 
\begin{align*}
K_{t_0}   = M_{\text{yes}, t_0}   H   M_{\text{yes},t_0}^{-1}+       \frac{\<\psi_{t_0}|[H,M_{\text{yes},t_0}^\dagger M_{\text{yes},t_0}]|\psi_{t_0}\>}{2\|M_{\text{yes}}|\psi_{t_0}\>\|^2} \, ,
\end{align*} along with the observation that  $\<\psi_{t_0}|[H,M_{\text{yes},t_0}^\dagger M_{\text{yes},t_0}]|\psi_{t_0}\>=  0$.
As a result,  the probabilistic QFI  at $t_0$, defined as the QFI associated to the states $\{  |\phi_t\>\}$,  satisfies  
\begin{align*}
Q^{\text{prob}}_{t_0}  &   =          \< \phi_{t_0} |   K^{^\dag }_{t_0}  K_{t_0}    |\phi_{t_0}\>    -    \< \phi_{t_0} |     K_{t_0}    |\phi_{t_0}\>^2   \\
&  = \frac{Q_{t_0}   }{ \epsilon^2}
  \end{align*}
This equation means that, whenever the original QFI is non-zero at a specific value $t_0$, the probabilistic QFI  can be made arbitrarily large with a suitable choice of $\epsilon$.   This  fact, combined with the CRB, seems to suggest that one can have unlimited precision.  However,  such a conclusion is an artifact of the pointwise character of  the CRB: the  probabilistic estimation scheme shown above has  unlimited precision only in an infinitesimally small neighborhood of  $t_0$.    In light of this observation, for  probabilistic metrology it is more sensible to choose the filter $M_{\text{yes}}$ independent of $t$, to assign a prior  $p(t)$ to the unknown parameter, and to minimize the expected variance.  
 Note that, since the estimate is produced only when the system passes through the filter, the expectation  has to be computed with respect  to the conditional probability distribution $p(t|\text{yes})  \propto \|  M_{\text{yes}}  |\psi_t\> \|^2  p(t)$.    
 The expected  variance, denoted by $ V^{\rm prob}_p     :  =   \int     {\rm d}  t  ~  p(t|{\rm yes})     ~       V_t $, 
is lower bounded by the inverse of the expected QFI, denoted by   $Q^{\rm prob}_p    :=\int   { \rm d t} ~   p(t|{\rm yes})~ Q_t$. In formula: 
\begin{align}\label{avecrao}
V^{\rm prob}_p    \ge    \frac1{  Q^{\rm prob}_p} \, .
\end{align} 
The proof of  equation (\ref{avecrao})  is an elementary application of  the pointwise quantum Cram\'er-Rao bound and of the Schwartz inequality:
\begin{align*}
  1   & \le  \left[   \int    {\rm d }t ~   p(t|{\rm yes})~    \sqrt{   V_t ~   Q_t  }   \right]^2\\
  &  \le  \left[ \int    {\rm d }  t ~   p(t|{\rm yes})~    V_t \right]  ~   \left[\int    {\rm d}  t  ~   p(t|{\rm yes})~  Q_t     \right]\\
    &   \equiv     V^{\rm prob}_p    ~ Q^{\rm prob}_p  \,  .
    \end{align*}
Note that  the average CRB of equation (\ref{avecrao}) is just a lower bound and the equality may not be achievable.    However, achievability is of no consequence for our arguments about the replication rates. 


\medskip 

\noindent {\bf Acknowledgments.}   This work is supported by the National Basic Research Program of China (973) 2011CBA00300 (2011CBA00301) and by the National Natural Science Foundation of China through Grants  11350110207, 61033001,  and 61061130540 and by the 1000 Youth Fellowship Program of China.
We are grateful to the two anonymous Referees for comments that stimulated a substantial strengthening of the original manuscript.
A particular acknowledgment goes to Xinhui Yang for drawing Figure 1. 

\medskip 

\noindent {\bf Author contributions.}  All authors contributed extensively to the work presented in this paper. 

\medskip 
\noindent{\bf  Competing financial interests. } The authors declare no competing financial interests.

\begin{widetext}

 \section*{Supplementary Note 1} 
 Here we lift the assumption  that the trace distance has to vanish as $1/M^{\beta}$, $\beta \ge 1$ in the proof of the SQL.  In addition, the argument here provides  a strong converse of the SQL and a proof that the probability of success of a replication process with rate $\alpha =  1  +\epsilon$ must be vanish faster than $\exp[- N^{\delta}]$ for every $\delta  < \epsilon$. 
 We prove these results for  states of the form $ |\psi_t\>  =   e^{-it H}  |\psi\>$, where $H$ is a suitable Hamiltonian on a finite dimensional Hilbert space. 
 Let us expand the input state as   
\begin{align}\label{decoo} 
 |\psi\>  =  \sum_{E\in\Spec (H)}   \sqrt{p_E}   |E\> \, , 
 \end{align} 
 where   $\Spec (H)$ denotes the spectrum of $H$ and $ |E\>\in\spc H$ is a normalized eigenvector corresponding to the eigenvalue $E $.  Note that even if $H$ is degenerate, we can always choose  the eigenvectors $|E\>$ so that all the eigenvalues in the  expansion  (\ref{decoo}) are distinct.    We  assume without loss of generality that the vectors $\{  |E\>  ~|~  E\in\Spec (H), p_E \not  = 0 \}$ span the whole Hilbert space $\spc H$ (if this is not the case, we can always restrict the attention to a suitable subspace).    In other words, we assume that $p_E  \not = 0$ for every $E$ in the spectrum of $H$: the number of eigenvalues $E$ such that $p_E \not = 0$, denoted by $K$, is then equal to the total number of eigenvalues, $|\Spec(H)|$.   When $N$ copies are given, we write   
 \begin{align}\label{psin}
  |\psi\>^{\otimes N}   =    \sum_{E\in\Spec \left(H^{(N)}\right)}   \sqrt{ p_{N,E}}   |N,E\>,
  \end{align}
  where $|N,E\> \in\spc H^{\otimes N}$ is a normalized eigenvector  of the $N$-particle Hamiltonian $H^{(N)}  :=  \sum_{j=1}^N  H_j$,   $  H_j$ denoting the operator $H$ acting on  the Hilbert space of the  $j$-th particle.  
  
\subsection{Constraints from covariance.}
 Following the Methods, we describe  a generic probabilistic cloning  process as a
 covariant CP map  $\map P_{{\rm yes}}$,  satisfying   
  $ U_t^{\otimes M}    \map P_{{\rm yes}} (\rho)    U_t^{\otimes  M  \dag}   =  \map P_{{\rm yes}}  \left(     U_t^{\otimes N} \rho   U_{t}^{\otimes  N\dag}\right)$  
   for every $t$ and for every state $\rho$, where  $ U_t:  =  e^{-it  H}$.  
Moreover, we decompose the probabilistic  process $\map P_{{\rm yes}}$ into $  \map P_{{\rm yes}} =  \map   C_{{\rm yes}} (  M_{{\rm yes}}  \cdot M^\dag_{{\rm yes}})$,   where   $\left[M_{{\rm yes}},    U_t^{\otimes N}    \right]=0$ for every $t$. Note that, by construction,  the quantum channel $\map C_{{\rm yes}}$ must be covariant, namely  
\begin{align}\label{cicov}
   U_t^{\otimes M}    \map C_{{\rm yes}} (\rho)    U_t^{\otimes  M  \dag}   =  \map C_{{\rm yes}}  \left(     U_t^{\otimes N} \rho   U_{t}^{\otimes  N\dag}\right)   \end{align}
for every $t$ and for every $\rho$.  It is convenient to put the constraints  on $M_{{\rm yes}}$ and $\map C_{{\rm yes}}$ in an explicit form.        Expressing $ U_t^{\otimes N} $ 
as  $ U_t^{\otimes N}   =  e^{-it  H^{(N)}}$  
 we can rewrite the condition  $\left[M_{{\rm yes}},    U_t^{\otimes N}    \right]=0 \, \forall t$ as 
\begin{align}\label{myes}
 M_{{\rm yes}}  =   \sum_{E\in\Spec \left( H^{(N)}\right)}    \pi_E  ~ |E,N\>\<E,N|~, 
\end{align}
for some coefficients $\pi_E  \in  [0,1]$.    
For  $\map C_{{\rm yes}}$, we express the constraint in terms of the Choi operator  \cite{choi}, denoted by $  C_{{\rm yes}}$, which is an operator on the tensor product Hilbert space $\spc H^{\otimes M}  \otimes \spc H^{\otimes    N}$.     Recall that the action of  $\map C_{{\rm yes}}$ can be expressed as \cite{lopresti}
\begin{align}\label{mauro}
\map C_{{\rm yes}} (\rho) =  \Tr_{N}  [   (I_M  \otimes \rho^T)   C_{{\rm yes}}] 
\end{align}   
where $\Tr_N$ is the partial trace over the $N$ input spaces, $I_M$ is the identity over the $M$ output spaces, and $\rho^T$ is the transpose of $\rho$.  
Using this fact, the  symmetry condition of equation (\ref{cicov}) becomes     $[ C_{{\rm yes}},  U_{t}^{\otimes M}  \otimes U_{-t}^{\otimes N}]  =  0$  for every $t$.  
This commutation relation implies that   $C_{{\rm yes}}$ can be written in the block diagonal form 
\begin{align}\label{s-channel-1}
&C_{{\rm yes}}=\bigoplus_{\mu  \in  \Spec  \left(  H^{(M)}  - H^{(N)}  \right) }   C_{\mu},\\
\nonumber &C_{\mu} =\sum'_{E, E' \in \Spec \left(  H^{(N)}\right)} [C_\mu]_{E,E'} ~  \left |   M,   E   +  \mu     \right \>\!  \left\<   M,    E' + \mu    \right|\otimes  \left |N, E\right\>\!\left\<N,  E'\right| , 
\end{align} 
where  the summation $\sum'$ is restricted to the values of $E$ ($E'$) such that $E +  \mu$    ($E'  +  \mu$) belongs to $\Spec (H^{(M)})$.    
 
\subsection{ Upper bound on the fidelity.}
Using  equations (\ref{cicov}) and  (\ref{mauro}),   we express the cloning fidelity   as  
\begin{align}
\nonumber F^{{\rm prob}}_{\rm{wc}}  [   M_{{\rm yes}} ,   C_{{\rm yes}}]  &:=  \inf_{t}         \< \psi_t|^{\otimes M}   \map C_{{\rm yes}}   \left(   |\Phi_t^N\>\<  \Phi_t^N| \right)      |\psi_t\>^{\otimes M}    \qquad \qquad |\Phi_t^N\>   =    \frac{M_{{\rm yes}}  |\psi_t\>^{\otimes N}  }{  \|  M_{{\rm yes}}  |\psi_t\>^{\otimes N}  \|^2}   \\
 \label{fidelity2} &  =  \frac{  \<  \psi |^{\otimes M }   \<  \psi|^{\otimes N}    C_{{\rm yes} }    |  \psi\>^{\otimes M }      |  \psi\>^{\otimes N }}{ \| M_{{\rm yes}} |  \psi\>^{\otimes N} \|^2   }  \, .
\end{align}
Now the goal is to find the maximum of the fidelity in equation (\ref{fidelity2}) over  all operators $M_{{\rm yes}}$,  subject to equation (\ref{myes}), and over all Choi operators $C_{{\rm yes}}$,  subject to equation (\ref{s-channel-1}) and to the trace-preserving condition   $\Tr[\map C_{{\rm yes}}(\rho)]  =  \Tr[\rho] $ for every state $\rho$.   In terms of the Choi operator, the preservation of the trace can be expressed as \cite{lopresti}
\begin{align*}
\Tr_M[   C_{{\rm yes}}]  =    I_N \, .
\end{align*}
where $\Tr_M$ denotes the partial trace over the $M$ output spaces and $I_N$ denotes the identity matrix  on the subspace   containing the states $|\psi_t\>^{\otimes N}$.  
For $C_{{\rm yes}}$ as in equation (\ref{s-channel-1}),    the above   condition  can be expressed  as 
\begin{align}\label{tp}
\sum'_{\mu   \in  \Spec  \left(  H^{(M)}  - H^{(N)}  \right) }   [C_\mu]_{E,E} =1  \qquad \forall E\in   \Spec \left(  H^{(N)}\right),
\end{align}
where the summation is restricted to the values of $\mu$  such that $ E  +\mu $ belongs to $\Spec \left( H^{\otimes M}\right)$.   Denoting by   $\|  H\|_{\infty}$   the operator norm $\|  H\|_{\infty}  :  =  \max_{E\in\Spec (H) }  |E|$, we have the following  
\begin{lemma}[Upper bound on the fidelity]\label{lem:fbound}
Let $M_{{\rm yes}} $ be a filter operator satisfying equation (\ref{myes}) and let  $C_{{\rm yes}} $ be a Choi operator satisfying Eqs. (\ref{s-channel-1}) and (\ref{tp}).  Then, for every $E_{\delta}  :  0\le E_\delta  \le   N  \|     H \|_{\infty}$, the fidelity $F^{\rm prob}_{\rm wc}  [   M_{{\rm yes}} ,   C_{{\rm yes}}]$  is upper bounded as   
\begin{align}\label{fbound}
F^{{\rm prob}}_{\rm{wc}} [M_{{\rm yes}}, C_{{\rm yes}}] \le    \left[ \sqrt{  \max_{\mu}  \left( \sum'_{|E |  \le E_\delta   }  p_{M,E+\mu}  \right)}          +  \sqrt{    \frac{  \max_{|E| >  E_{\delta}} ~ p_{E,N}~    }{   p_{{\rm yes}}}           (N+1)^{K} } \right]^2 \, ,
\end{align}
where $K$ is the number of distinct eigenvalues of $H$. 
\end{lemma}
 
\medskip

{\bf Proof.} Using equations (\ref{psin}) and (\ref{myes}), we  express the states $|\psi\>^{\otimes M} $  and $M_{{\rm yes}}|\psi\>^{\otimes N}$ and the probability $  p_{{\rm yes}}  $ as 
\begin{align*}
|\psi\>^{\otimes M}   &=  \sum_{E  \in \Spec \left(  H^{(M)}\right)}  \sqrt{p_{M,E}}    |  M, E \>  \\
 M_{{\rm yes} }   |\psi\>^{\otimes N}     &  =  \sum_{E\in \Spec \left(  H^{(N)}\right)}    \pi_E   \sqrt{ p_{N,E} } ~      |N, E  \>\\
  p_{{\rm yes}}&=  \|  M_{{\rm yes}}|\psi\>^{\otimes N} \|^2   =    \sum_{E\in \Spec \left(  H^{(N)}\right)}    \pi^2_E   ~ p_{N,E}  .   
\end{align*} 
Inserting these expressions into equation (\ref{fidelity2}) and using equation (\ref{s-channel-1}), we get
  \begin{align}
  F_{{\rm yes}}^{{\rm prob}} [M_{{\rm yes}},C_{{\rm yes}}]=\sum_{\mu} \left(\sum'_{E,E'\in\Spec \left (  H^{(N)} \right)}    [C_\mu]_{E,E'}    ~    \pi_E  \pi_{E'}    \sqrt{\frac{p_{M,  E  +\mu } p_{N,E}}{p_{{\rm yes}}}  }        \sqrt  {\frac {p_{M,  E'  +\mu  } p_{N,E'}}{p_{{\rm yes}}}  }   \right). 
\end{align}
Now, the positivity of the operator $C_{{\rm yes}}$ (and  hence of the matrices $C_{\mu}$)  implies   the inequality   $|[C_\mu]_{E,E'}  |\leqslant\sqrt{    [C_\mu]_{E,E}  [C_\mu]_{E',E'}}$.   Defining  $c_E^\mu :  =  [C_\mu]_{E,E} $,  
we then have
\begin{align}
\nonumber 
F_{{\rm yes}}^{{\rm prob}}  [M_{{\rm yes}}, C_{{\rm yes}}]&\le \sum_{\mu}\left(\sum'_{E\in \Spec\left(  H^{(N)}\right)}  
\pi_E \sqrt{\frac{c_E^\mu~ p_{M, E+ \mu}  p_{N,E}}{p_{{\rm yes}}}}  \right)^2\\
\nonumber
&=    \sum_{\mu   }\left(\sum'_{|E| \le E_
\delta   }  \pi_E   \sqrt{\frac{c_E^\mu p_{M,E+\mu}  p_{N,E}}{p_{{\rm yes}}}}    +          \sum'_{|E|  > E_\delta  }   \pi_E   \sqrt{\frac{c_E^\mu p_{M,E+\mu}  p_{N,E}}{p_{{\rm yes}}}}          \right)^2\\  
   & \le   \left[\sqrt{ \sum_{\mu }   \left(\sum'_{|E |  \le E_\delta   }    \pi_E   \sqrt{\frac{c_E^\mu p_{M,E+\mu}  p_{N,E}}{p_{{\rm yes}}}}       \right)^2} +  \sqrt{ \sum_{\mu }  \left(    \sum'_{E  > E_\delta  }   \pi_E   \sqrt{\frac{c_E^\mu p_{M,E+\mu}  p_{N,E}}{p_{{\rm yes}}}}   \right)^2 }   \right]^2  \, ,  \label{finale}
\end{align}
 having used the triangular inequality.  
The first term in equation (\ref{finale}) is upper bounded as  
\begin{align*}     \sum_{\mu} \left(  \sum'_{|E| \le E_\delta  }   \pi_E  \sqrt{ \frac{ c_E^\mu  p_{M,E+\mu}  p_{N,E}}{p_{{\rm yes}}}} \right)^2      
&    \le       \sum_{\mu }   \left(  \sum'_{|E| \le E_\delta  }    p_{M, E+ \mu }     \right)       \left(    \sum'_{|E|  \le E_\delta } {\frac{  \pi_E^2  c_E^\mu p_{N,E}   }{p_{{\rm yes}}}} \right)  \\
&    \le    \max_{\mu}  \left(  \sum'_{|E| \le E_\delta  }    p_{M, E+ \mu }     \right)                     ~   \left (  \sum_{\mu}                  \sum'_{|E| \le E_\delta } {\frac{  \pi_E^2  c_E^\mu p_{N,E}   }{p_{{\rm yes}}}} \right) \\
&    \le    \max_{\mu}  \left(  \sum'_{|E| \le E_\delta  }    p_{M, E+ \mu }     \right).
 \end{align*} 
The last inequality comes from the observation \begin{align*}
\sum_{\mu}  \sum_{ |E|  \le E_{\delta}}'   \pi_E^2  c_E^{\mu}   p_{N,E}  =  \sum_{ |E|  \le E_{\delta}}   \sum_{\mu}'   \pi_E^2  c_E^{\mu}   p_{N,E}     = \sum_{ |E|  \le E_{\delta}}   \pi_E^2   p_{N,E}   \le         \sum_{ E\in\Spec\left(   H^{(N)}\right) }   \pi_E^2   p_{N,E}     =  p_{{\rm yes}},
\end{align*}
having used the trace-preserving condition $\sum_\mu'   c_E^\mu  =1  \, \forall E\in \Spec \left(H^{(N)}\right)$  [cf. equation (\ref{tp})].  The second term in equation (\ref{finale})  can be upper bounded as  
\begin{align*}
    \sum_{\mu }  \left(    \sum'_{E  > E_\delta  }   \pi_E   \sqrt{\frac{c_E^\mu p_{M,E+\mu}  p_{N,E}}{p_{{\rm yes}}}}   \right)^2        &  \le 
   \frac{  \max_{E>  E_\delta} \left( p_{N,E}\right) }{   p_{{\rm yes}}}   
  \sum_{\mu }  \left(    \sum'_{|E|  >  E_{\delta}  }      \sqrt{c_E^\mu p_{M,E+\mu}  }   \right)^2  \\  
& \le \frac{  \max_{E>E_\delta} \left( p_{N,E}\right) }{   p_{{\rm yes}}}      \sum_{\mu }    \left(    \sum'_{E  > E_{\delta}  }        c_E^\mu      \right)     \left(    \sum'_{E >  E_{\delta}  }     p_{M,E+\mu}   \right) \\
& \le \frac{  \max_{E>E_\delta} \left( p_{N,E}\right) }{   p_{{\rm yes}}}      \sum_{\mu }    \left(    \sum'_{E  > E_{\delta}  }        c_E^\mu      \right)   \\
  &\le    \frac{  \max_{E >  E_{\delta}} \left( p_{N,E}\right) }{   p_{{\rm yes}}}       \left(    \sum'_{E >  E_{\delta}  }     1     \right) \\
  &\le    \frac{  \max_{E >  E_{\delta}} \left( p_{N,E}\right) }{   p_{{\rm yes}}}       \left(    \sum'_{E \in \Spec\left(   H^{(N)}\right)  }     1     \right) 
  \end{align*}
Note that the eigenvalues of  $ H^{(N)} $  are all of the form $E ({\mathbf n})  =  \sum_{E\in\Spec (H)}     n_E   E$, where $ \mathbf n  :  =  ( n_E)_{E\in\Spec (H) } $ is a partition of $N$ into $ K$ nonnegative integers.  This fact implies that  the number of distinct eigenvalues of $ H^{(N)} $  is upper bounded by the number of partitions, given by $ {N+K-1\choose K-1} $.    Hence, we can continue the above chain of inequalities with 
\begin{align*}  
    \sum_{\mu }  \left(    \sum'_{E  > E_\delta  }   \pi_E   \sqrt{\frac{c_E^\mu p_{M,E+\mu}  p_{N,E}}{p_{{\rm yes}}}}   \right)^2         &  \le  \frac{  \max_{E >  E_{\delta}} \left( p_{N,E}\right) }{   p_{{\rm yes}}}          {N+K-1\choose K-1} \\
   &  \le  \frac{  \max_{E >  E_{\delta}} \left( p_{N,E}\right) }{   p_{{\rm yes}}}        (N+1)^{ K} 
      \end{align*}
$\blacksquare$

\subsection{Strong converse  of the SQL}
We are now ready to prove that the probability of success of a replication process with rate $\alpha =  1  +\epsilon$ must be vanish faster than $\exp[- N^{\delta}]$ for every $\delta  < \epsilon$:  
\begin{theo}[Strong converse of the SQL for information replication]\label{theo:strongSQL}
Let $(\map P_{{\rm yes}, N})_{N  \in\mathbb N}$ be a sequence of replication processes producing $M  \ge c  N^{  1+  \epsilon}$ replicas, with $0<  \epsilon\le 1$ and $c>0$. If the  success probability of the $N$-th process, denoted by $p_{{\rm yes}, N}$,  satisfies  
\begin{align}\label{bastaa}
\liminf_{N \to \infty} ~  \frac{   p_{{\rm yes}, N} }{e^{-N^\delta}}  >  0
\end{align}
for some $\delta  < \epsilon$, then the fidelity of the replicas   vanishes in the limit $N\to\infty$.
  \end{theo}

\medskip 

{\bf Proof.}   The proof consists in applying lemma \ref{lem:fbound} with    $  E_\delta  :  =  \sqrt{ 2 N^{1+\delta}}  \|  H  \|_{\infty}$. 
Our goal is  to   show that, under the assumption of equation (\ref{bastaa}),   the r.h.s. of equation (\ref{fbound}) vanishes in the large $N$ limit.   In order to prove this fact, we assume without loss of generality that $\< \psi  |  H  |\psi\>  = 0$ (this condition can be always enforced by re-defining $H'  :=    H-\<  \psi|  H  |\psi\>$).  With this choice,  the first term in the r.h.s. of equation (\ref{fbound}) satisfies
\begin{align*}   \lim_{N \to \infty}   \max_{\mu}     \left(  \sum'_{|E| \le E_\delta  }    p_{M, E+ \mu }     \right)  
&   =    \lim_{N \to \infty}    \max_{\mu }   \left(  \int_{-E_\delta  }^{+  E_\delta}\d E ~  \frac{e^{-(E+\mu)^2/ (2 M   \< H^2\>) }}{\sqrt{2\pi  M \< H^2\>}}      \right) \\
&    \le    \lim_{N \to \infty}   \left(  \int_{-E_\delta  }^{+  E_\delta} \d E~     \frac{e^{-E ^2/ (2 M   \< H^2\>) }}{\sqrt{2\pi  M \< H^2\>}}      \right)                    ~   \\  
&\le \lim_{N\to \infty}  \frac{2 E_\delta} {\sqrt{2\pi  M \< H^2\>}}\\
&  \le  \lim_{N \to \infty}  \frac  {  2\|  H\|_\infty} { \sqrt{\pi  c N^{\epsilon- \delta}  \< H^2\>} } \\
&  =0 \, .
  \end{align*} 
In the first equality we used the central limit theorem, which allows one to approximate the cumulative distribution of $p_{M, E}$ with that of the Gaussian  $  \frac{e^{-E^2/ (2 M   \< H^2\>) }}{\sqrt{2\pi  M \< H^2\>}}$ up to an error of order $1/\sqrt M$   \cite{berry}.   For the second  term in the r.h.s. of equation (\ref{fbound}), we note the bound 
\begin{align*}
    \max_{|E| >  E_{\delta} }  p_{N,E} &    \le   {\rm prob}   (|E| >  E_{\delta} )  \\
    &
     \le   2e^{-    E^2_\delta/ (N \|  H\|_\infty^2)   } 
\end{align*}
following from  Hoeffding's inequality \cite{hoeffding}. Using this bound,  the second  term in the r.h.s. of equation (\ref{fbound}) is upper bounded as
  \begin{align*}
\lim_{N\to\infty}   \frac{  \max_{|E| >  E_{\delta} }  (p_{N,E})      }{   p_{{\rm yes},N}} ~ (N+1)^{ K}  &    \le  \lim_{N\to\infty}   \frac{ 2 e^{-    E^2_\delta/ (N \|  H\|_\infty^2)   } }{   p_{{\rm yes},N}} ~ (N+1)^{ K}\\   
&    \le  \lim_{N\to\infty}   \frac{  2e^{-  2 N^{\delta}   } }{   p_{{\rm yes},N}} ~ (N+1)^{ K}\\
&    \le \frac{   \lim_{N\to \infty}  2 e^{-N^\delta}  (N+1)^{ K} }   {   \liminf_{N\to\infty}   {   p_{{\rm yes},N}}/{  e^{-   N^{\delta}   } }}  \\
& = 0 \,  .     
\end{align*}
$\blacksquare$

\section*{Supplementary Note 2}

Here we establish the Heisenberg limit for the probabilistic estimation of the parameter $t$  from the clock state $|\psi_t\>  =  e^{-it H}  |\psi\>  \in\spc H$. 
Conditionally on the successful outcome, the joint state of the systems emerging from the filter is  
\begin{equation}\label{s-probe}
|\phi_t\>=\frac{M_{{\rm yes}}|\psi_t\>}{\|M_{{\rm yes}}|\psi_t\>\|} 
\end{equation}
Let $V^{{\rm prob}}_t $ be the variance in the estimation of $t$ from the state $|\phi_t\>$.  By the Cram\'er-Rao bound, the variance is bounded by the inverse of the Quantum Fisher Information (QFI), given by \cite{helstrom,QFImetric} 
\begin{align}
Q^{{\rm prob}}_t   
 \label{qfi} &    = 4  \left[  \<  \phi_t |   \left(H^{{\rm prob}}_t\right)^2  |\phi_t\>    -  \<  \phi_t |   H^{{\rm prob}}_t  |\phi_t\>^2  \right]   
\end{align} 
where  $H^{{\rm prob}}_t$  is the Hamiltonian defined by   the conditions $i\frac{\d}{\d t}|\phi_t\>=H^{{\rm prob}}_t |\phi_t\>$ and   $H^{{\rm prob}~ \dag}_t  =  H^{{\rm prob}}_t  $.
Note that, in principle, $H^{{\rm prob}}_t$ can depend on $t$, even if the original Hamiltonian $H$ does not.   By explicit calculation we obtain  $H^{{\rm prob}}_t  |\phi_t\>  =   K_t    |\phi_t\>$, where $K_t$ is the non-Hermitian operator
\begin{equation}\label{s-H}
K_t  :  = \left(M_{{\rm yes}}   H   M_{{\rm yes}}^{-1}+   \frac {  \d  M_{{\rm yes}}}{\d t} M_{{\rm yes}}^{-1}  +     \frac{\<\psi_t|[H,M_{{\rm yes}}^\dagger M_{{\rm yes}}]|\psi_t\>}{2\|M_{{\rm yes}}|\psi_t\>\|^2}   -  \frac {i  \<  \psi_t  |    \frac{  \d}{\d t} (  M_{{\rm yes}}^\dag M_{{\rm yes}})    |\psi_t\>}{2  \|   M_{{\rm yes}}  |\psi_t\>  \|^2} \right) \, .
\end{equation}
Since $H^{{\rm prob}}_t  |\phi_t\>  =   K_t    |\phi_t\>$, substituting into equation   (\ref{qfi}) gives   
\begin{align}\label{prob-fisher}
Q_t^{{\rm prob}}   =      4 \left(      \< \phi_t |   K^{^\dag }_t  K_t    |\phi_t\>    -     \< \phi_t |     K_t    |\phi_t\>^2   \right).
\end{align} 
Note that in Eqs. (\ref{s-H}) and (\ref{prob-fisher}) we allowed the choice of the filter $M_{{\rm yes}}$ to depend on $t$, considering the most general case. 
However, as  discussed in the Methods Section, assuming that $M_{{\rm yes}}$ can be chosen to be  $t$-dependent is unrealistic. 
 For this reason, we impose that $\d M_{{\rm yes}}/\d t =  0$,  getting
\begin{equation}\label{s-HH}
K_t    = \left(M_{{\rm yes}}   H   M_{{\rm yes}}^{-1}+        \frac{\<\psi_t|[H,M_{{\rm yes}}^\dagger M_{{\rm yes}}]|\psi_t\>}{2\|M_{{\rm yes}}|\psi_t\>\|^2}    \right) \, .
\end{equation}
Hence, the quantum Cram\'er-Rao bound gives the inequality 
\begin{equation}
V_t^{{\rm prob}}  \ge \frac1{ 4 \left ( \<  \phi_t| K_t^\dag  K_t  |\phi_t\>  -   \<  \phi_t|  K_t  |\phi_t\>^2 \right)}  \, .  
\end{equation}

\section*{Supplementary Note 3}

Here we show that the average of the variance with respect to the conditional probability distribution  
\begin{align}
p(t|  {\rm yes})  =  \frac{ p(t) \<  \phi_t|  M_{\rm yes}^\dag  M_{\rm yes}  |\phi_t\>}{  \int  \d \tau   \,  p(\tau) \<  \phi_\tau|  M_{\rm yes}^\dag  M_{\rm yes}  |\phi_\tau\>}
\end{align} 
is lower bounded as  
\begin{align}
V_p^{\rm prob}  \ge \frac 1 {  (E_{\max}  -  E_{\min})^2} 
\end{align}     
in the case of periodic evolutions of period $T$ with uniform prior $p(t)  =  1/T$. Let us start from a general result valid for arbitrary priors $p(t)$.  
Using the Cram\'er-Rao bound and the  Schwartz inequality, the average variance can be lower bounded  as 
\begin{align}\label{crave}
V^{{\rm prob}}_{p}    \ge    \frac1{  Q_p^{{\rm prob}}  } \, .   
\end{align}
where $  Q_p^{{\rm prob}}$  is the inverse of the average QFI 
\begin{align}
  Q_p^{{\rm prob}}     =       \int   {\rm  d} t  ~   p(t|{\rm yes})~  Q_t^{{\rm prob}}  ~.      
\end{align}
In turn, the average QFI can be upper bounded as follows 
 \begin{lemma}[Upper bound on the average QFI]\label{lem:twirl}
 The average QFI for the clock states $\{  |\psi_t\>  =  e^{-it H}  |\psi\>  \}$ 
 is upper bounded as  
 \begin{align}\label{fishbound}
 Q_p^{{\rm prob}}    \le     4 \sup'_{P\ge 0}
    \left\{  \frac{    ~    \<  \psi|     ~H~     \map T_p( P)       ~ H~ |\psi  \> }{    \<  \psi |    \map T_p(P)    |\psi \>    }-  \left[  \frac{  {\rm Re}   \left(  \<  \psi  | ~ \map T_p (  P)  H  ~ |\psi\>\right)}{  \< \psi|  \map T_p (P)  |\psi\>}\right]^2 \right\} 
\end{align}
where 
$\map T_p$ is the twirling map
$\map T_p  (P) :  =        \int   {{ \rm d}  \tau }     ~ p(\tau) ~    e^{i\tau  H }    P   e^{-i\tau  H }$ and $\sup'$ denotes the supremum over the operators $P$ such that the denominators are non-zero.

 \end{lemma}  
  
\noindent {\bf Proof.}   
 We start  from the definition of $K_{t}$ for a $t$-independent filter operator [cf. equation (\ref{s-HH})]
\begin{align*}
K_t    = M_{{\rm yes}}   H   M_{{\rm yes}}^{-1}+      \frac{\<\psi_t|[H,M_{{\rm yes}}^\dagger M_{{\rm yes}}]|\psi_t\>}{2\|M_{{\rm yes}}|\psi_t\>\|^2}  
    \end{align*} 
and note that 
 \begin{align*}
K_{t}  - \< \psi_t|  K_t |\psi_t\>      =   M_{{\rm yes}}  H M_{{\rm yes}}^{-1}  -  \frac{  \< \psi_t|  M_{{\rm yes}}^\dag  M_{{\rm yes}}  H  |\psi_t\>}{ \|  M_{{\rm yes}} |\psi_t\>  \|^2}  \, .  
\end{align*}
Then, using equation (\ref{prob-fisher})  we obtain 
\begin{align*}
Q^{{\rm prob}}_t &    =      4\left(    \< \phi_t |   K^{^\dag }_t  K_t    |\phi_t\>    -      \< \phi_t |     K_t    |\phi_t\>^2  \right)\\  
  &  =  4\left[    \< \phi_t |     ( K^{\dag}_t   -  \< \psi_t|  K_t |\psi_t\> ) ( K_t  -  \< \psi_t|  K_t |\psi_t\> )     |\phi_t\> \right] \\      
   &  =  4  \left[ \frac  {  \< \psi_t|   ~  H  M_{{\rm yes}}^\dag M_{{\rm yes}}  H~  |\psi_t\>}  {    \| M_{{\rm yes}}  |\psi_t\>  \|^2}    -  \left|   \frac  {  \< \psi_t|   ~    M_{{\rm yes}}^\dag M_{{\rm yes}}  H~  |\psi_t\>}  {   \| M_{{\rm yes}}  |\psi_t\> \|^2  }        \right|^2  \right]      \\
  &  \le  4  \left\{ \frac  {  \< \psi_t|   ~  H  M_{{\rm yes}}^\dag M_{{\rm yes}}  H~  |\psi_t\>}  {  \|  M_{{\rm yes}}  |\psi_t\> \|^2 }    - \left[  \frac  {  {\rm Re}   \left(    \< \psi_t|   ~    M_{{\rm yes}}^\dag M_{{\rm yes}}  H~  |\psi_t\>\right)}  {    \| M_{{\rm yes}}  |\psi_t\> \|^2 }        \right]^2  \right\}  \, .  \end{align*} 
Averaging over the probability distribution $p(t |{\rm yes})$, we then have 
\begin{align*}
\nonumber      Q_p^{{\rm prob}}   & =    \int   {\rm d} t  ~   p(t|{\rm yes})~  Q_t^{{\rm prob}}\\
 & \le  4  \int        {\rm d} t   ~     p(t|{\rm yes})          \left\{   \frac  {  \< \psi_t|   ~  H  M_{{\rm yes}}^\dag M_{{\rm yes}}  H~  |\psi_t\>}  {  \<  \psi_t|  M_{{\rm yes}}^\dag M_{{\rm yes}}  |\psi_t\>  }    - \left[  \frac  {     {\rm Re} \left( \< \psi_t|   ~    M_{{\rm yes}}^\dag M_{{\rm yes}}  H~  |\psi_t\>\right)}  {  \<  \psi_t|  M_{{\rm yes}}^\dag M_{{\rm yes}}  |\psi_t\>  }        \right]^2  \right\}     \\
& \le    4  \left[     \int   {\rm d} t   ~     p(t|{\rm yes})  ~         \frac  {  \< \psi_t|   ~  H  M_{{\rm yes}}^\dag M_{{\rm yes}}  H~  |\psi_t\>}  {  \<  \psi_t|  M_{{\rm yes}}^\dag M_{{\rm yes}}  |\psi_t\>  }\right]    -    4  \left[  \int   {\rm d} t   ~     p(t|{\rm yes})   ~   \frac  {   {\rm Re}\left(  \< \psi_t|   ~    M_{{\rm yes}}^\dag M_{{\rm yes}}  H~  |\psi_t\>\right)}  {  \<  \psi_t|  M_{{\rm yes}}^\dag M_{{\rm yes}}  |\psi_t\>  }        \right]^2       \\
&  =  4  \left[   \frac{  \int   {\rm d} t   ~     \< \psi_t|   ~  H  M_{{\rm yes}}^\dag M_{{\rm yes}}  H~  |\psi_t\>}  { \int {\rm d} \tau ~    \<  \psi_\tau|  M_{{\rm yes}}^\dag M_{{\rm yes}}  |\psi_\tau\>  }\right]    -    4  \left[  \frac  {  {\rm Re}   \left(  \int   {\rm d} t   ~     \< \psi_t|   ~    M_{{\rm yes}}^\dag M_{{\rm yes}}  H~  |\psi_t\> \right)}{ \int {\rm d} \tau~  \<  \psi_\tau |  M_{{\rm yes}}^\dag M_{{\rm yes}}  |\psi_\tau \>  }        \right]^2       \\
&  =     4  \left[  \frac  {  \< \psi|   ~  H  \map T_p(M_{{\rm yes}}^\dag M_{{\rm yes}} )   H~  |\psi\>}  {  \<  \psi|  \map T_p(M_{{\rm yes}}^\dag M_{{\rm yes}})  |\psi\>  }  \right]  -  4 \left[       \frac  {   {\rm Re}   \left(\< \psi|   ~    \map T_p(M_{{\rm yes}}^\dag M_{{\rm yes}})  H~  |\psi\> \right)}  {  \<  \psi|  \map T_p(M_{{\rm yes}}^\dag M_{{\rm yes}})  |\psi\>  }        \right]^2       \\
\nonumber 
& \le     4   \sup'_{P  \ge 0}       
\left\{
  \frac{ \<  \psi|     ~H~     \map T_p( P)       ~ H~ |\psi  \> }{    \<  \psi |    \map T_p(P)    |\psi \>    } -
  \left[  \frac{ {\rm Re}\left( \<  \psi  | ~ \map T_p (  P)  H  ~ |\psi\>\right)}{  \< \psi|  \map T_p (P)  |\psi\>}\right]^2  \right\} \, . 
 \end{align*}
 $\blacksquare$ 

Using the above bound in the case of uniform prior $p(t)= 1/T$ it is immediate to provide  probabilistic  Heisenberg limit for clock states.    Indeed, averaging an arbitrary  operator $P$ over the uniform prior gives an operator $\map T_p(P)$  that satisfies  the commutation relation
\begin{align*}
[\map T_p(P), e^{-i t H }]= 0 \quad \forall t \, .
\end{align*}       
Defining $|\phi_P \>  :  =  \sqrt{ \map T_p(P)  }  |\psi\>/  \|  \sqrt{ \map T_p(P)  }  |\psi\> \|$, equation (\ref{fishbound}) becomes 
 \begin{align*}
 Q_p^{{\rm prob}}   \le   4  \sup'_{P \ge 0 }       \left(  \<  \phi_P |   H^2  |  \phi_P \>    -   \<  \phi_P |   H  |  \phi_P \>^2 \right  ).
 \end{align*}
 Maximizing the variance of $H$ over all operators $P$, we then get the upper bound $Q_p^{{\rm prob}}  \le  ( E_{\max}-  E_{\min})^2$.    Finally, the average quantum Cram\'er-Rao bound of equation (\ref{crave})  yields  $V^{{\rm prob}}_{p}  \ge   1/  ( E_{\max}-  E_{\min})^2$.

\section*{Supplementary Note 4}
  
Here we consider  non-periodic evolutions  and show that  for every $\epsilon >0$  there exists a probability distribution  $p (t)$  such that, for every filter $M_{{\rm yes}}$,    
\begin{align}\label{epsilon}
 V_{p}^{{\rm prob}} \ge \frac1 {         ( E_{\max} -  E_{\min} )^2 +  \epsilon  }.
\end{align}
The argument here is the same used in Supplementary Note 3, only with the technical complication that now the uniform distribution has to be approximated with a Gaussian of large variance.  Let us denote   the Gaussian by
$  p_\sigma (t)  =  \sqrt{ 1/2\pi\sigma^2}  e^{-t^2/(2 \sigma^2)} $ and, for a generic operator $A$, let us denote by $s(A)$ and $m(A)$  the values $  s(A)  =  \<  \psi|  HAH  |\psi\>/\<\psi|A|\psi\>$ and $m(A)  =    {\rm Re} \left[   \<  \psi|  AH  |\psi\>/\<\psi|A|\psi\> \right]$, respectively.   With this notation,   the upper bound of lemma \ref{lem:twirl} can be expressed as 
 \begin{align}\label{fishbound1}
 Q^{{\rm prob}}_{p_\sigma}    \le    4 \sup^{\prime}_{P\ge 0}
     \left\{   s \left[   \map T_{p_\sigma}( P) \right]      -         m \left[ \map T_{p_\sigma} (  P)  \right]^2 \right\} 
\end{align} 
Now, for every fixed operator $P\ge 0$, one has
\begin{align*}
\map  T_{p_\sigma} (P)     &=     \int  \frac {{ \rm d}  \tau }   {\sqrt{2 \pi \sigma^2}}   ~ e^{  -  \frac{  \tau^2}{2\sigma^2}} ~    e^{i\tau  H}    P   e^{-i\tau  H}   \\
&  =  \sum_{E, E'  \in  \Spec  (H )}   ~  e^{-\frac{  \sigma^2 (E- E')^2}2}   \<  E  |  P  |E'\> ~   |E \>  \<   E' |\\
&      =  {\rm diag}(P) +   \delta P 
\end{align*}
where 
\begin{align*}
    {\rm diag} (P)  & : =  \sum_{E\in\Spec (H)}     \<  E  |  P  |E\>  ~|E\>\< E|   \\
     \delta P   &: =    \sum_{ E\not = E'}   ~  e^{-\frac{  \sigma^2 (E- E')^2}2}   \<  E  |  P  |E'\> ~   | E \>  \< E' |       .  
\end{align*}    
Hence, for large variance $\sigma^2$, the operator $\map T_{p_\sigma}  (P)$ can be made arbitrarily close to ${\rm diag} (P)$.  Clearly, if in equation (\ref{fishbound1})  we can replace  $\map T_{p_\sigma}  (P)$   with    ${\rm diag} (P)$ up to an error $\epsilon/4$, then the result is proved:  indeed, ${\rm diag} (P)$ commutes with $e^{-itH}$, and therefore, defining the state $  |\phi_P\> :  =  {\rm diag} (P)  |\psi\>  / \|   {\rm diag} (P)  |\psi\>   \| $ we have 
\begin{align*}
 s  \left[ {\rm diag }(P) \right]  -      m \left[ {\rm diag }(P) \right]^2    & =   \<  \phi_P|  H^2  |\phi_P\>  -  \<  \phi_P|  H  |\phi_p\>^2 \\ 
      & \le  (E_{\max}  -  E_{\min})^2/4    \, ,
\end{align*}
which implies  $Q^{{\rm prob}}_{p_\sigma}     \le  (E_{\max}  -  E_{\min})^2  +  \epsilon$ and   $V_{p}^{{\rm prob}} \ge \left[         ( E_{\max} -  E_{\min} )^2 +  \epsilon  \right]^{-1}$.  The rest of the proof consists just in showing that it is indeed possible to replace $\map T_{p_\sigma}  (P)$ with ${\rm \diag }(P)$ in equation (\ref{fishbound1}), up to a small error that vanishes uniformly in $P$ when the variance $\sigma^2$ is large. 
To this purpose, note that the first term in the r.h.s. of equation (\ref{fishbound1}) satisfies
 \begin{align}\label{continuami}
s  \left[  \map T_{p_\sigma}( P)  \right] 
 & \le      \frac{      \<  \psi |     ~H~      {\rm diag}(P)       ~ H~ |\psi  \>  +   |      \<  \psi |     H      ~\delta P  ~      H |\psi  \>|  }{\left |    \<  \psi |  ~ {\rm diag} ( P)   ~  | \psi \> -    |   \<  \psi |         \delta P    |\psi  \>|   \right |} \, . 
\end{align}
Here,  $|   \<  \psi |         \delta P   |\psi  \>|$ can be upper bounded as 
\begin{align*}
 |   \<  \psi |         \delta P    |\psi  \>|  &\le  \sum_{E\not = E'} \left |  \sqrt{  p_E p_{E'}}  ~ e^{-\frac{  \sigma^2 (E- E')^2}2}   \<  E  |  P  |E'\> \right|  \\
 &  \le   e^{-\frac{  \sigma^2 \Delta_{\min}^2}2}   ~     \left(\sum_{E\in\Spec(H)}   \sqrt{  p_E  \<  E  |   P  |E\>  }\right)^2  \qquad \qquad \Delta_{\min}  :  =   \min_{ E\not =  E'}    |  E-E'|
\\
 & \le   e^{-\frac{  \sigma^2 \Delta_{\min}^2}2}   ~      \sum_{E \in \Spec(H)} \<  E  |   P  |E\>  \\
 &  \le     e^{-\frac{  \sigma^2 \Delta_{\min}^2}2}   ~   K    ~  \|  {\rm diag} (P) \|_{\infty} \\
  &  = \delta ~  \|  {\rm diag} (P) \|_{\infty}    \qquad \qquad \delta  :  =       e^{-\frac{  \sigma^2 \Delta_{\min}^2}2}   ~   K~.   
   \end{align*}
Similarly,  $   |      \<  \psi |     H     \delta P        H |\psi  \>| $  can be upper bounded as   
\begin{align*}
   |      \<  \psi |     H     \delta P        H |\psi  \>|  & \le    \sum_{E,E'}  \left|   E~ E'~  \sqrt{  p_E p_{E'} }   ~ e^{-\frac{  \sigma^2 (E- E')^2}2}   \<  E  |  P  |E'\> \right|  \\
&\le  \|  H\|_{\infty}^2~      \sum_{E,E'} \left |  \sqrt{  p_E p_{E'}}  ~ e^{-\frac{  \sigma^2 (E- E')^2}2}   \<  E  |  P  |E'\> \right|  \\
&\le  \|  H\|_{\infty}^2~   \delta   ~  \|  {\rm diag} (P) \|_{\infty} \, .
\end{align*}
Inserting the above bounds in equation (\ref{continuami}), we  obtain
\begin{align}
\nonumber    
s  \left[  \map T_{p_\sigma}( P)  \right]   
 &  \le      \frac{      \<  \psi|     ~H~      {\rm diag}(P)       ~ H~ |\psi  \>  +       \delta      \|  {\rm diag}  (P)  \|_{\infty} ~   \|  H\|_\infty^2 }{  | \<  \psi |   {\rm diag} ( P)     |\psi \> -         \delta      \|  {\rm diag}  (P)  \|_{\infty} |} 
\\ \label{continuamiancora} 
&    =  \frac{      \<  \psi|     H      D_P        H |\psi  \>  +       \delta      ~   \|  H\|_\infty^2 }{  | \<  \psi |   D_P  |\psi \> -         \delta   |   }   \qquad  D_P  : =\frac {  {\rm diag} (P)}{  \|  {\rm diag} (P)\|}_{\infty} \, .  
\end{align}
Note that, for every $P\ge 0$   ($P\not =  0$), one has 
\begin{align*}
\<\psi  |D_P  |\psi\>  =  \sum_{E}  p_E~    \frac{ \<  E |  P    |E\>}{\| {\rm diag}(P) \|_{\infty}}  \ge   p_{\min}  \, .
\end{align*} 
Hence, choosing $\delta  <  p_{\min}$  (i.e. choosing the variance $\sigma^2$ in the Gaussian prior to be sufficiently large), we can remove the modulus in the denominator of equation (\ref{continuamiancora}), thus getting    
\begin{align*}
s  \left[  \map T_{p_\sigma}( P)  \right]  &  \le      \frac{      \<  \psi|     H      D_P        H |\psi  \>  +       \delta      ~   \|  H\|_\infty^2 }{   \<  \psi |   D_P  |\psi \> -         \delta      }  \\
& \le   \frac{      \<  \psi|     H      D_P        H |\psi  \>   }{   \<  \psi |   D_P  |\psi \>   (1-         \delta/    \<  \psi |   D_P  |\psi \>  )   }   + \frac{       \delta      ~   \|  H\|_\infty^2}  { \<  \psi |   D_P  |\psi \> -         \delta      }   \\
&   \le        \frac 1 {(1 -         \delta/p_{\min}   )}      ~   s  \left(    D_P\right) +         \frac{\delta       \|  H\|_\infty^2}{  p_{\min}  - \delta}
\end{align*}
Note that for sufficiently small $\delta$ (i.e. for sufficiently large variance $\sigma^2$), the upper bound can be arbitrarily close to $ s(     D_P)  $.  In other words, for every desired $\epsilon_1 >  0$, we can find a $\delta>  0$ such that   
\begin{align}\label{epsilonuno}
  s  \left[      \map T_{p_\sigma}( P) \right]   & \le   s(D_P)    + \epsilon_1 ,
\end{align}
for every possible $P$.  Regarding the second term in equation (\ref{fishbound1}), it satisfies 
\begin{align*}
       \left|   m\left [    \map T_{p_\sigma}  (P)\right] \right|     &  \ge    \left|   \frac  {  | \< \psi |  ~ {\rm diag} (P)  H  |\psi\>  | -      | {\rm Re}   ( \< \psi |  ~ \delta P  H  |\psi\>) | }   {   \< \psi |  {\rm diag} (P)   |\psi \>  + |\<\psi | \delta P |\psi\>|}     \right|
\end{align*}  
where $|   \<  \psi |         \delta P  H   |\psi  \>|$ can be upper bounded as 
\begin{align*}
 |   \<  \psi |         \delta P ~  H   |\psi  \>|  &\le  \sum_{E,E'} \left |  \sqrt{   p_E p_{E'}} ~   E'  ~ e^{-\frac{  \sigma^2 (E- E')^2}2}      \<  E  |  P  |E'\> \right|  \\
 &  \le   \|H\|_{\infty}      \sum_{E,E'} \left |  \sqrt{   p_E p_{E'}}  ~ e^{-\frac{  \sigma^2 (E- E')^2}2}   \<  E  |  P  |E'\> \right|  \\
 &\le  \|  H\|_\infty    ~ \delta ~ \| {\rm diag}(P)\|_{\infty} \, . 
   \end{align*}
Hence,  we have 
\begin{align*}
          \left|   m\left [    \map T_{p_\sigma}  (P)\right] \right|       
            &   \ge     \left|   \frac  {   |\< \psi |  ~ {\rm diag} (P)  H  |\psi\> |   -\delta  \|  H\|_{\infty}   \|  {\rm diag} (P) \|_{\infty}  }   {   \< \psi |  {\rm diag} (P )  |\psi \>  + \delta   \|  {\rm diag} (P)  \|_{\infty}}  \right| \\
              &  \equiv       \left|   \frac  { |  \< \psi |  ~   D_P  H  |\psi\>  |  -\delta  \|  H\|_{\infty}      }   {   \< \psi |  D_P   |\psi \>  + \delta  }  \right|           
               \, .
  \end{align*}    
Now, assuming without loss of generality that the eigenvalues of $H$ are all positive, we have the  inequality $  \<  \psi|  D_P H |\psi\>  \ge   p_{\min}  E_{\min}$.  Hence, choosing $\delta  <  p_{\min}  E_{\min}/\|  H\|_{\infty}$ we obtain 
   \begin{align*}   
         \left|   m\left [    \map T_{p_\sigma}  (P)\right] \right|       
            &  \ge     \frac  {   \< \psi |  ~ D_P   H  |\psi\>    -\delta  \|  H\|_{\infty}    }   {   \< \psi |  D_P   |\psi \>  + \delta  }  \\
&    \ge    \frac 1 {(1 +         \delta/p_{\min}   )}          ~            m\left  (  D_P  \right)       -     \frac{\delta \|  H\|_{\infty}}{  p_{\min}+\delta}  \, .    
\end{align*} 
Again, note that for sufficiently small $\delta$ (i.e. for sufficiently large variance $\sigma^2$), the lower bound can be arbitrarily close to $             m\left  (  D_P  \right) $. In other words,  for every $\epsilon_2>0$  one can find a $\delta >0$ such that 
 \begin{align}\label{epsilondue}   m\left[  \map T_{p_\sigma}  (P)\right]  &   \ge     m(D_P)       - \epsilon_2 \, ,
\end{align}
for every possible $P$.  Combining the inequalities (\ref{epsilonuno}) and (\ref{epsilondue}) and suitably choosing $\epsilon_1$ and $\epsilon_2$,  we have that for every $\epsilon>0$ there exists a $\delta >0$ such that  
\begin{align*}
  \sup'_{P\ge 0}
         \left|   m\left[   \map T_\sigma (P) \right] \right| &   \le      \sup'_{P\ge 0}  m(D_P)   +  \epsilon/4.       
\end{align*}  
Since by definition $  m(   D_P)  =  m\left[   {\rm diag}  (P) \right]$, this concludes the proof that in equation (\ref{fishbound1}) one can substitute $\map T_{p_\sigma}  (P)$ with ${\rm diag} (P)$ up to an error $\epsilon$.      $\blacksquare$

\section*{Supplementary Note 5} 

Here we provide a strong converse of the HL, showing that every process that replicates quantum information beyond the HL must have vanishing fidelity. 
\begin{theo}[Strong converse of the HL for information replication]
Let $(\map P_{{\rm yes}, N})_{N  \in\mathbb N}$ be a sequence of replication processes producing  $M  \ge c  N^{  2+\epsilon}$ replicas, with $\epsilon  >  0$ and $c> 0$.   
Then, the fidelity of the replicas vanishes in the limit $N\to \infty$. 
  \end{theo} 

{\bf Proof.}  The proof consists in applying lemma \ref{lem:fbound} of Supplementary Note 1  with  $  E_\delta  :  =  N  \|  H \|_{\infty} $.  
With this choice, all the eigenvalues of $H^{(N)}$ satisfy $E  \le E_{\delta}$.      
Hence, equation (\ref{fbound})  contains only the first term.  
Again, we assume without loss of generality that $\< \psi  |  H  |\psi\>  = 0$.  With this choice we have      
\begin{align*}   \lim_{N \to \infty}   \max_{\mu}     \left(  \sum'_{|E| \le E_\delta  }    p_{M, E+ \mu }     \right)  
&   =    \lim_{N \to \infty}    \max_{\mu }   \left(  \int_{-E_\delta  }^{+  E_\delta}\d E ~  \frac{e^{-(E+\mu)^2/ (2 M   \< H^2\>) }}{\sqrt{2\pi  M \< H^2\>}}      \right) \\
&    \le    \lim_{N \to \infty}   \left(  \int_{-E_\delta  }^{+  E_\delta} \d E~     \frac{e^{-E ^2/ (2 M   \< H^2\>) }}{\sqrt{2\pi  M \< H^2\>}}      \right)                    ~   \\  
&    \le   \lim_{N \to \infty}     \frac{  2 E_\delta} {\sqrt {2 \pi  M \<  H^2\>}} \\       
&    \le   \lim_{N \to \infty}     \frac{  2    \|  H\|_{\infty}} {\sqrt {2 \pi c N^{\epsilon} \<  H^2\>}} \\       
&  =0.
  \end{align*} 
  In the first equality we used the central limit theorem,  approximating  the cumulative distribution of $p_{M, E}$ with that of the Gaussian  $  \frac{e^{-E^2/ (2 M   \< H^2\>) }}{\sqrt{2\pi  M \< H^2\>}}$ up to an error of order $1/\sqrt M$   \cite{berry}. 
$\blacksquare$

\section*{Supplementary Note 6}

Here we show that the spectrum of $H^{(M)}$ contains the spectrum of $H^{(N)}$ up to a translation by a suitable amount $\delta E_0 \approx  (M-N)  \<  \psi|  H |\psi\>$.   In order to to this,   recall that the eigenvalues of $H^{(N)}$ are parametrized by partitions of $N$ into $K $ non-negative integers, as  
\begin{align}\label{mapping}
E (\mathbf n)  &=  \sum_{E\in\Spec (H)}    n_E ~   E \, . 
 \end{align}
Let us denote by $\map P_{N,K}$ the set of partitions of $N$ into $K$ non-negative integers and  choose a partition   $\mathbf n_0 \in \map P_{N,K}$  such that the frequencies   $n_{0,E}/N$ are close to the corresponding probabilities $p_E$, i.e.
\begin{align*}
|  n_{0,E}  -   N p_E |  \le 1    \qquad  \qquad \forall E \in\Spec (H) .
\end{align*}
Similarly, let us choose a partition   $\mathbf m_0 \in \map P_{M,K}$  such that 
\begin{align*}
|  m_{0,E}  -   M p_E |  \le 1    \qquad  \qquad \forall E \in\Spec (H) .
\end{align*}
Now, for every partition $\mathbf n \in \map P_{N,K}$ one can define the partition $\mathbf m\in\map P_{M,K}$ as  
\begin{align*}
\mathbf m  =  \mathbf n -\mathbf n_0   +  \mathbf m_0 \, . 
\end{align*}  
Hence, for every eigenvalue $E$ of $H^{(N)}$ there is an eigenvalue $E'$ of $H^{(M)}$ of the form 
\begin{align*}
E'   =  E  +  \delta E_0  \qquad \qquad \delta E_0  : =  E (\mathbf m_0)  -  E (\mathbf n_0) \, . 
\end{align*} 
Note that, by definition of ${\bf n}_0$ and ${\bf m}_0$, $\delta E_0$ is close to $(M-N)  \<  \psi  |  H  |\psi\>$, in the sense that the difference between these two values does not grow with $N$.   
Precisely, one has 
\begin{align*}
\left |  \delta E_0  -  (M-N)  \<  \psi  |  H  |\psi\> \right|&   =      \left|  \sum_{E\in\Spec (H)}  E ~  \left[  m_{0,E}   -   n_{0,E}  -  (M-N)  p_E    \right]  \right|\\
&  \le   \|  H \|_{\infty}     ~      \sum_{E\in\Spec(H)}   \left(  |   m_{0,E}  -  Mp_E  |    +  |   n_{0,E}  -  N  p_E       |\right)\\
&\le      \|  H \|_{\infty}    ~  2 K   ~,
\end{align*}
having used the fact that, by definition,  $  |   m_{0,E}  -  Mp_E  | \le 1  $  and    $ |   n_{0,E}  -  N  p_E   |\le 1$ for all the $K$  values of $E$.
 \section*{Supplementary Note 7}\label{subsect:filt}

Here we show that the fidelity of our filter is lower bounded as
\begin{align}\label{hoeff}
F^{{\rm prob}}_{\rm{wc}}  [  M_{{\rm yes}}, \map C_{{\rm yes}}] \ge       1  -   2 K  \exp\left[  \frac{-  2    N^2 p^2_{\min}} M    +     \frac{4N}{  K M}\right] \, .
\end{align}  
whenever   $N\ge   1/p_{\min}$,  ~   $p_{\min}  := \min _E  p_E$.    
By definition, the fidelity is given by
 \begin{align}
\nonumber F_{\rm{wc}}^{{\rm prob}} [ M_{{\rm yes}}, \map C_{{\rm yes}} ]  &:=  \inf_{t}     | \left  \<  \psi_t |^{\otimes M}   V  |\Phi^N_t\right \> |^2  \qquad    \left |\Phi_t^{N}\right\> = :  \frac{  M_{{\rm yes}}  |  \psi_t\>^{\otimes N}  }{ \|    M_{{\rm yes}}| \psi_t \>^{\otimes N}  \| }\\
\label{fidelityfinale} &  =       \sum_{E  \in\Spec (H^{(N)})}   p_{M,E+ \delta E_0 } \, .
 \end{align}
 In order to lower bound the fidelity, it is useful to express the probabilities $p_{M,E}$ in terms of the partitions of $M$ that parametrize the eigenvalues of $H^{(M)}$.   Let us expand the state $ |\psi\>^{\otimes M}$ on the  basis given by  the symmetrized states  
\begin{align}\label{symme}
|M, \mathbf m \>  \propto     \sum_{\pi \in S_M}   U_\pi     \left(  \bigotimes_{E\in\Spec (H)}   | E\>^{\otimes m_E}  \right)   \qquad \qquad {\bf m}  \in  \map P_{M,K}\, ,  
\end{align}
 where $\pi\in S_M$ is a permutation  and $U_\pi$ is the operator that permutes the $M$ Hilbert spaces according to $\pi$.  With this definition, we have
 $ |\psi\>^{\otimes M}   =   \sum_{\mathbf{m} \in\map P_{M,K}}   \sqrt {  q_M (\mathbf{ m})}   ~  |M,\mathbf m\>  ,$
 where  $q_M (\mathbf m)$ is
 the multinomial distribution  
  \begin{align}\label{multi}
q_M  (\bf m)       & :=     M!   \prod_{E\in \Spec (H) }       \frac{ p_E^{m_E}}{m_E!} \,  .
\end{align} 
Now, since the state $ |M,{\bf m}\>$ is an eigenvector of $H^{(M)}$ with eigenvalue $E({\bf m})  =  \sum_{E\in \Spec (H)}  m_E E$, the probability that the $M$ particles   have total energy $E$ is  given by
\begin{align*}
p_{M,E}    & =    \sum_{
\begin{array}{c} 
{\bf m} \in\map P_{M,K}  \, , \\
E({\bf m})   = E
\end{array}}
   q_M (\bf m) 
  \end{align*}
and equation (\ref{fidelityfinale}) becomes 
  \begin{align*}
\nonumber F_{\rm{wc}}^{{\rm prob}} [ M_{{\rm yes}}, \map C_{{\rm yes}} ]  & =\sum_{E  \in  \Spec \left(  H^{(N)}\right)}    \sum_{
\begin{array}{c}
{\bf  m} \in\map P_{M,K}  \\    E({\bf m})  =  E +  \delta E_0
\end{array}
}   q_M (\bf m)  \\
& =\sum_{{\bf n}  \in  \map P_{N,K}}    \sum_{
\begin{array}{c}
{\bf  m} \in\map P_{M,K}  \\    E({\bf m})  =  E  (  {\bf n })   +   E ( {\bf   m}_0 ) -  E(  {\bf n_0}  ) 
\end{array}
}   q_M (\bf m)    \\
& \ge \sum_{{\bf n}  \in  \map P_{N,K}}    q_M  (  {\bf n }   +  {\bf m_0}  -{\bf   n}_0 ) 
 \, . 
 \end{align*}
having used the definition  $\delta E_0: =   E({\bf m}_0) -  E  (\bf {n}_0)$ and  the fact that the eigenvalues of $H^{(N)}$ are parametrized by partitions in $\map P_{N,K}$. 
Since for a generic partition  $\mathbf n \in  \map P_{N,K}$,  the difference $   n_E   - n_{0,E}   $ can assume all integer values in the interval $I_{E,N}  =  [-  p_E  N, (1-p_E)  N ]$, we have 
  \begin{align*}
F_{\rm{wc}}^{{\rm prob}} [ M_{{\rm yes}}, \map C_{{\rm yes}} ] 
&  \ge   \sum_{
\begin{array}{c}
\mathbf m  \in \map P_{M,K} \\
   \forall E  :    m_E  -  m_{0,E}\in  I_{E,N}    
  \end{array}}  ~  q_M(\mathbf m)    
  \\
&  \ge  1  -  \sum_{E\in \Spec (H)}  \sum_{
\begin{array}{c}
\mathbf m  \in \map P_{M,K} \\
  |  m_E  -  m_{0,E}  |   > N  p_{E,\min}   \\
   p_{E,\min} : =  \min \{  p_E,  1-p_E\}   
  \end{array}}   q_M (\mathbf m)   \,  .
  \end{align*}
Recall that, for every $E$, the frequency $m_{0,E}/M$ is close to the probability $p_E$:  precisely,  $  | m_{0,E}  -   p_E  M  | \le 1  $.  
  Hence, choosing  $N$ such that $   N  p_{E,\min}    \ge  1 $ for every $E$,  we obtain 
 \begin{align*}
F_{\rm{wc}}^{{\rm prob}} [ M_{{\rm yes}}, \map C_{{\rm yes}} ]   &  \ge  1  -  \sum_{E\in \Spec(H)}  \sum_{
\begin{array}{c}
\mathbf m  \in \map P_M \\
  |  m_E  -  p_E   M |   > N  p_{E,\min}   -1   
  \end{array}}   q_M (\mathbf m)  \\
&  \ge 1  -   \sum_{E  \in \Spec (H)}       2  \exp  \left[  - \frac{ 2    (N  p_{E,\min}     -1)^2}{M}   \right]\, ,
\end{align*}
having used Hoeffding's inequality  \cite{hoeffding} for the last bound.  Taking the worst-case over $E$ we then obtain 
\begin{align*}
F_{\rm{wc}}^{{\rm prob}} [ M_{{\rm yes}}, \map C_{{\rm yes}} ]   &  \ge  1  -  2  K   \exp  \left[  - \frac{ 2      (N  p_{\min}     -  1 )^2}{  M}     \right]  \qquad  \qquad p_{\min}  :  = \min_{E}  p_E\\
&  \ge  1  -  2  K    \exp  \left[  - \frac{ 2     N^2 p^2_{\min}   } {  M}    +  \frac{4   N  p_{\min}}   M     \right]  \\  
&  \ge  1  -  2  K  \exp  \left[  - \frac{ 2     N^2 p^2_{\min}  } {  M}    +  \frac{4   N}{K M }    \right]\, ,
\end{align*}
having used the relation $p_{\min}  \le 1/K$.  
$\blacksquare$

\section*{Supplementary Note 8}

Here we evaluate the probability of success of the filter 
\begin{align}\label{nostrofiltro}
M_{\rm yes}:=\sum_{E\in\Spec(H^{(N)})}\pi_E |N,E\>\!\<N,E|\qquad \pi_E=\gamma\sqrt{\frac{p_{M,E+\delta E}}{p_{N,E}}}
\end{align} 
where $\gamma$ is a suitable constant, showing that it decays exponentially with $N$ for every replication rate $\alpha> 1$.  Using equation (\ref{nostrofiltro}), one can express the probability of the successful outcome as   
\begin{align*}
p_{{\rm yes}}  &   =   \| M_{{\rm yes}}|\psi\>^{\otimes N}  \|^2  \\
   &  =    \sum_{ E  \in  \Spec (H^{(N)}) }  \pi_E^2  ~  p_{N, E}  \\
  &  =  \gamma^2    \sum_{ E  \in  \Spec (H^{(N)}) }     p_{M, E+  \delta E_0}\\
  &\le \gamma^2,
\end{align*}   
where the value of $\gamma$ is constrained by the requirement that each  $\pi_E$ in equation (\ref{nostrofiltro}) be smaller than $1$. Due to the definition $\pi_E  :  = \gamma  \sqrt{ p_{M,E +\delta E_0}/  p_{N,E}}$,  this condition implies the upper bound
\begin{align*}
\gamma^2  \le \frac{p_{N,E}}{p_{M,E+  \delta E_0}} \qquad \forall E\in\Spec \left(  H^{(N)}\right) . 
\end{align*}   
In particular, choosing $E  =  N  E_*$, where $E_*$ is the eigenvalue of $H$ with maximum modulus, we get  
\begin{align*}
\gamma^2  \le \frac{p_{N,N  E_*}}{p_{M,N E_*+  \delta E_0}} \, .
\end{align*}   
Let us now express the probabilities  $p_{N,N  E_*}$ and $p_{M,N E_*+  \delta E_0}$ in terms of partitions of $N$ and $M$, respectively.  
Since $E_*$ has maximum modulus, there is only one partition of $N$ such that $E({\bf n})  =  N E_*$, namely  the partition ${\bf n}_*$ where all the entries are null, except to the one corresponding to $E_*$, i.e. $n_{*,E}  =   N ~ \delta_{E , E_*}$. Hence, we have  $p_{N, N E_*}  \equiv   q_N  ({\bf n}_*)$  and
\begin{align*}
\gamma^2     &\le     \frac{  q_{N,   {\bf  n}_*} }{ \sum_{   
{\bf m} :
     E({\bf  m})  =  N E_*  +  \delta E_0  }  \qquad
 q_{M,  {\bf m}  }}    \\
  &\le     \frac{  q_{N,   {\bf  n}_*} }{   q_{M,  {  \bf m}_*   }}         \qquad \qquad  {\bf m}_*  :  =   {\bf n}_*  -  {\bf n}_0 +  {\bf m}_0 \, . 
  \end{align*}
  Inserting the expression for the multinomial distributions [equation (\ref{multi})], we then get
  \begin{align*}
   \gamma^2  &  \le       \frac 1 {   M! }   \prod_E     (   m_{*,E})!    ~     p_E^{    n_{0,E}  -  m_{0,E}  }  ~ \\
  &  \approx       \frac 1 {   M! }   \prod_E     (   m_{*,E})!   ~     p_E^{   (N-M  ) p_E   }     \\
  &  =     \left[  \frac 1 {   M! }   \prod_E        (   m_{*,E})!   \right]   ~       e^{  (M-N)  H({\bf p})}     \\
  & \approx             \left\{ \frac 1 {   M! }   \prod_E        \left[  n_{*,E}  +  (M-N)  p_E\right]! \right\}   ~    e^{  (M-N)  H({\bf p})} ~
   \end{align*}  
  where in the second and  fourth lines we approximated $n_{0,E}$  ($m_{0,E}$) with $p_E N$  ( $p_E M$), while in the  third line we introduced the notation $H({\bf p})  :=  -\sum_E  p_E  \ln  p_E$  for the Shannon entropy of the vector ${\bf p}  : = (p_E)$. In the second and fourth lines,  $\approx$ denotes asymptotic equivalence up to polynomial factors, that is $  f(N)  \approx  g(N)$ if there are two polynomials $\rm{poly}_1 (N)$ and ${\rm poly}_2 (N)$ such that, for sufficiently large $N$,  $ f(N)    \ge    {\rm poly}_1  (N)   g(N) $   [denoted by $f (N)  \gtrsim  g(N) $]  and $   f(N)   \le  {\rm poly}_2  (N)  g(N)$  [denoted by $  f(N)  \lesssim g(N)$]. 
Using Stirling's approximation, we then get
\begin{align*}
 \gamma^2   \lesssim  e^{    -M  H({\bf f}_*)}   ~  e^{ (M-N)   H({\bf p})}    
 \qquad    {\bf f}_*   :  =  \frac NM  {\bf p}_* +  \left(  1  - \frac N M \right)  {\bf p} \, ,   \qquad {\bf p}_*  : =  \frac{ {\bf n}_*} N\,  .    
\end{align*} 
 Note that,  the quantity $N/M$ vanishes in the large $N$ limit  (by hypothesis the replication rate is larger than $1$). Hence, by Taylor-expanding to the second order in $N/M$, we get 
 \begin{align*}  
 H({\bf f}_*)  &    =          \left\{   H({\bf p})   -  \frac NM  ~   \left[  \sum_{  E}     (p_{*,E}  -  p_E)  \ln p_E \right]   +  O\left( {N^2}/{ M^2}\right)        \right\}  \\
 &  =     \left\{   \left(  1  - \frac N M \right)     H({\bf p})   - \left(  \frac N M \right)       \ln p_{E_*}    +  O(N^2/M^2)        \right\}    \, ,
\end{align*}   
having used the condition $p_{*,E}  =  \delta_{E,E_*}$.   In conclusion, we obtained  
 \begin{align}\nonumber
\gamma^2 &\lesssim  \exp  \left[          N      \ln p_{E_*}     +  O\left( \frac  {N^2} M \right)  \right] 
 \end{align}
\medskip 
which implies the exponential scaling of the success probability:
\begin{align*}
p_{{\rm yes}}    &  \le \gamma^2 \lesssim      \exp  \left\{      -    N       \left[  \ln \left(  \frac 1 { p_{E^*}}  \right)  +  O  \left( \frac NM\right)  \right]  \right\}    
\end{align*} 
The above equation proves that the probability of success decreases exponentially in $N$, with a constant asymptotically close to $\ln  \left( \frac 1 {p_{E^*}} \right)$.  




\section*{Supplementary Note 9}

We now show how the success probability of our filter can be increased to the maximum values allowed by theorem \ref{theo:strongSQL}:  for a super-replication process with rate $\alpha =  1 +\epsilon$, we show that the probability can assume any value that is negligible with respect to  $\exp[-N^{\epsilon}]$  in the large $N$ limit.  Precisely,   we devise  super-replication processes that produce $M$ copies, with $M$ bounded as
\begin{align}\label{Mbounds}
c_1  N^{1+\epsilon}  \le     M \le  c_2  N^{1  +\epsilon}
\end{align} 
for every desired $\epsilon  \in [0,1)$ and $c_1,c_2>0$. The feature of these super-replication processes is that the  fidelity approaches $1$ in the limit $N\to \infty$ and the  success probability  is lower bounded  as 
\begin{align}\label{boh}
p_{{\rm yes}}\gtrsim  e^{-  N^{\epsilon}   f(N)}  \, ,
\end{align} 
where $f(N)$ is any fixed positive function such that
  \begin{align}\label{properties1}
\lim_{  N  \to \infty  }   f(N)  &=  \infty \\
\label{properties2} \lim_{N\to\infty}  \frac{f(N)} {N^{1-\epsilon}}   &=  0  \, .
\end{align}  
 For the cloning process, we choose  the isometry  
\begin{align*}
V  =  \sum_{ {\bf n}  \in\map P_{N,K} }     |  M,  {\bf  n }  -  {\bf n}_0  +  {\bf m}_0  \>   \<  N,  {\bf n}|  ,
\end{align*}  
where the vectors $  |N,{\bf n}\>$ and $|M, {\bf m}\>$ are defined as in equation (\ref{symme}) of Supplementary Note 7.  
Then,  for a fixed function $f$ satisfying equation (\ref{properties1}) and (\ref{properties2}), we define the set of partitions 
\begin{align*}
\map P_{f} :=   \left \{  {\bf n}  \in  \map P_{N,K}  ~|~ \forall E  :  ~    | x_E |  \le   \sqrt{  \xi  M  f(N)}   \qquad x_E :  = n_E -  n_{0,E} \right\}  \, ,
\end{align*} 
where  $\xi>0$ is a constant, whose value will be  fixed later.    Finally, we define the filter operator as 
\begin{align*} 
 M_{{\rm yes}}=\sum_{\mathbf n \in  \map P_{f} } \pi_{\mathbf n}   |N, \mathbf n \>\!\<N, \mathbf n | , \qquad
\pi_{\mathbf n}  =  
\left\{    
\begin{array}{ll} 
\gamma \sqrt{    \frac {     q_{M,{\bf n}  - {\bf n}_0  +  {\bf m}_0 }   }    {  q_{N,\bf n}   }}  \qquad   &     {\bf n}\in \map  P_{f}    \\
 &\\
0  \qquad \qquad  &   {\bf n}  \not \in  \map P_{f}  
\end{array}
\right. 
\end{align*} 
where $\gamma$ is a suitable constant.  The combination of the filter $M_{{\rm yes}}$ with the isometric channel $\map C_{{\rm yes}}  (\rho ) = V \rho V^\dag$ gives a probabilistic process with fidelity 
\begin{align*}
\nonumber F^{{\rm prob}}_{\rm{wc}}  [ M_{{\rm yes}},\map C_{{\rm yes}} ]   &=    \frac{\left|  \< \psi|^{\otimes M}    V  M_{{\rm yes}}  |\psi\>^{\otimes N} \right|^2  }{  \|  M_{{\rm yes}}  |\psi\>^{\otimes N}  \>  \|^2} \\  
\nonumber &=    \sum_{ {\bf n}  \in  \map P_{f}}    q_{M,{\bf n}  -  {\bf n}_0  +  {\bf m}_0}  \\
&  \ge  1 -  \sum_{E\in\Spec(H)} \sum_{ 
\begin{array}{c}
{\bf    m  }  \in \map P_{M}  \\
\nonumber
|m_E  -   m_{0,E}|  >  \sqrt {\xi M  f(N)}\\  
\end{array}}  
    q_{M,{\bf m} }  \, .
\end{align*}
   Recalling that, by definition,  $|  m_{0,E}  -  M p_E |  \le 1$, we then have 
   \begin{align}
\nonumber  F^{{\rm prob}}_{\rm{wc}}  [ M_{{\rm yes}},\map C_{{\rm yes}} ] &  \ge  1 -  \sum_{E\in\Spec(H)} \sum_{ 
\begin{array}{c}
{\bf    m  }  \in \map P_{M}  \\
\nonumber
|m_E  -   M p_E |  >  \sqrt {\xi M  f(N)}  -1\\  
\end{array}}  
    q_{M,{\bf m} }  \\
\nonumber &   \le  1  -  2  \sum_{E\in \Spec (H)}    \exp\left[  -  \frac{ 2  \left(     \sqrt {\xi M  f(N)}  -1 \right)^2 }  M  \right]\\
\label{hoeff2}   &  \ge  1  -  2    K  \exp  \left[  -  2  \xi   f(N)    +  4          \sqrt {\frac{\xi f(N)}M}    \right]    ~,
\end{align} 
where the second to last bound comes from Hoeffding's inequality \cite{hoeffding}.    
 Since $f$ satisfies equation (\ref{properties2}),   equation (\ref{hoeff2})  implies that the fidelity tends to 1 as $\exp[-2  \xi f(N)]$ in the limit $N\to \infty$.

We now show that the probability of success of the cloning process has the best possible scaling with $N$ allowed by theorem \ref{theo:strongSQL}.  To this purpose, observe that $p_{{\rm yes}}$ satisfies  
\begin{align*}
p_{{\rm yes}}   &  =  \|  M_{{\rm yes}}  |\psi\>^{\otimes N}  \|^2  \\  
&  =   \sum_{ {\bf n}\in\map P_{f}  }     \pi_{\bf n}  ~  q_{N,{\bf  n}}    \\
 &   =    \gamma^2       \sum_{  {\bf n}\in\map P_{f}   }           q_{M, {\bf n}  -  {\bf n}_0  + {\bf m}_0 }   ,
 \end{align*}
 where the constant $\gamma$ can be chosen to be 
 $ \gamma  =   \min_{{\bf n}\in\map P_{f}}   \frac{  q_{N , {\bf n}}}{  q_{M , {\bf n}  -{\bf n}_0 +  {\bf m}_0}} $.
Now, due to equation (\ref{hoeff2}) one has  $p_{{\rm yes}} \approx   \gamma^2$ in the large $N$ limit.  Hence, we only need to estimate $\gamma$.  
To this purpose, note that 
\begin{align*}
 \frac{  q_{N , {\bf n}}}{  q_{M , {\bf n}  -{\bf n}_0 +  {\bf m}_0}}  &  = \frac  {  N! }{M!}  \prod_{E}  \left(   \frac{m_E!}{n_E!}   ~    p_E^{  n_{0,E}  -  m_{0,E} }  \right) \\
 &\approx         \left(     \frac  {  N! }{M!}  \prod_{E}    \frac{m_E!}{n_E!}   \right)  ~   e^{ (M-N)   H({\bf p})}   ~ \\
 &\approx        \left[     \prod_{E}  \frac{    \left(\frac{m_E}{  M } \right)^{m_E}   } {  \left(\frac{n_E}{  N } \right)^{n_E}  } \right]~   e^{ (M-N)   H({\bf p})}   ~ ,
\end{align*}
having approximated $n_{0,E}$   (respectively, $m_{0,E}$) with $p_E N$ (respectively, $ p_E M$) in the second line and having used the Stirling's approximation in the third line.  Expressing $n_E$ and $m_E$ as $n_E  =   n_{0,E}  +  x_E   $ and $m_E  =  m_{0,E}  +  x_E$,    we can approximate $n_{E}/N$  and $m_E/M$  with $p_E  +  x_E/N$ and $p_E  + x_E/M$, respectively,
thus obtaining
\begin{align}
\nonumber  \frac{  q_{N , {\bf n}}}{  q_{M , {\bf n}  -{\bf n}_0 +  {\bf m}_0}}     &\approx      \left[     \prod_{E}  \frac{  \left(        p_E    +  x_E/M\right)^{m_E}  }{    \left(  p_E  + x_E/N \right)^{n_E}   }  \right]    ~  e^{ (M-N)   H({\bf p})}     \\ 
\label{basta} &\approx                \exp\left\{   \sum_E  \left  [   ( M  p_E  +  x_E  )  \ln     \left(  p_E  +  \frac{x_E} M  \right)   -      ( N p_E  +  x_E  )  \ln     \left(  p_E  + \frac{ x_E}N  \right)       \right]\right\}      ~  e^{ (M-N)   H({\bf p})}    \, .  
 \end{align}
 Now, for every $\bf n$ in $\map P_f$, the deviation $  x_E    =   n_E  -  n_{0,E}  $  satisfies the conditions  
 \begin{align*}
 \lim_{N\to \infty} \frac{x_E}M     \le  \lim_{N\to\infty}   \sqrt{\frac{\xi  f(N)}  M}  = 0
 \end{align*}  and 
\begin{align*}   
\lim_{N\to \infty}   \frac{x_E}N  & \le   \lim_{N\to\infty}   \frac  {\sqrt{\xi     M  f(N)} } N  \\
     &\le   \lim_{N\to\infty}  \frac{  \sqrt{\xi      c_2   N^{1+\epsilon}    f(N)} } {N}\\
      &\le   \lim_{N\to\infty}   \sqrt{\frac{\xi     c_2     f(N)}  {N^{1-\epsilon}}} \\
      & =  0 \, ,
\end{align*} 
having used the bound $M  \le c_2  N^{1+\epsilon}$ and  equation (\ref{properties2}). Hence, for large $N$ both terms $x_E/M$ and $x_E/N$ are guaranteed to be small for large $N$ and we can  make a Taylor approximation in equation (\ref{basta}), thus getting
\begin{align}
\nonumber
 \frac{  q_{N , {\bf n}}}{  q_{M , {\bf n}  -{\bf n}_0 +  {\bf m}_0}}    &\approx   \exp\left [ -  \left( \frac{   \sum_E   x_E^2 }{  2p_E}  \right)   \left( \frac 1N  -  \frac 1 M\right)  \right] \\
 \nonumber & \ge     \exp\left[    \frac{ - K   \xi M   f(N) }{2p_{\min}}       \left( \frac 1N  - \frac 1M \right)\right] \\
 \label{ultrabasta} &  =   \exp\left[    \frac{ - K   \xi    f(N) }{2p_{\min}}       \left( \frac MN  -1 \right)\right]  \, .
 \end{align} 

Now, for  $ \epsilon =  0$  (that is, for $M  \le c_2  N,  \, c_2  >1$),  equation (\ref{ultrabasta}) gives  
\begin{align*}
 \frac{  q_{N , {\bf n}}}{  q_{M , {\bf n}  -{\bf n}_0 +  {\bf m}_0}}   &  \gtrsim     \exp\left[    \frac  {   - K       \xi      f(N) \left(c_2-1 \right)  }   {  2p_{\min}}   \right]   \,  .
 \end{align*} 
Choosing $\xi^{-1}  =    K \left(c_2 -1\right)    /(2p_{\min}) $ we obtain the desired scaling $  p_{{\rm yes}}  \gtrsim  e^{  -   f(N)}$, as promised by equation (\ref{boh}) with $\epsilon  = 0$.   Instead, for $\epsilon  >  0$, equation (\ref{ultrabasta}) gives 
\begin{align*}
 \frac{  q_{N , {\bf n}}}{  q_{M , {\bf n}  -{\bf n}_0 +  {\bf m}_0}}  
&  \approx       \exp  \left[       \frac  {  -   K (\xi-\omega)    f(N)}   {  2p_{\min}}  \left (\frac MN   \right)  \right] 
 \end{align*}
where $\omega  >0$ is an arbitrary constant compensating for the term $\exp [ O(  f(N))]$.   
Using the relation $M  \le c_2  N^{1+\epsilon}$, we then obtain 
\begin{align*}
 \frac{  q_{N , {\bf n}}}{  q_{M , {\bf n}  -{\bf n}_0 +  {\bf m}_0}}    & \gtrsim         \exp  \left[   - \frac{   K  (\xi  -\omega)    f(N)       c_2   N^{\epsilon} }{2 p_{\min}}     \right]  \,  .
 \end{align*}
Choosing $\xi $ and $\omega$ such that $(\xi-\omega )^{-1}   =  K   c_2 /(2p_{\min}) $, we obtain  the desired scaling   $p_{{\rm yes}} \gtrsim   e^{-  N^\epsilon  f(N)}$, as promised by equation (\ref{boh}).

\end{widetext}

\end{document}